# Musculoskeletal full-body models including a detailed thoracolumbar spine for children and adolescents aged 6 to 18 years

Stefan SCHMID[1,2,3,*], Katelyn A. BURKHART[1,4], Brett T. ALLAIRE[1], Daniel GRINDLE[1], Dennis E. ANDERSON[1,2]

[1]Beth Israel Deaconess Medical Center, Center for Advanced Orthopaedic Studies, Boston, MA
[2]Harvard Medical School, Department of Orthopaedic Surgery, Boston, MA
[3]Bern University of Applied Sciences, Department of Health Professions, Division of Physiotherapy, Bern, Switzerland
[4]Massachusetts Institute of Technology, Harvard-MIT Health Sciences and Technology Program, Cambridge, MA

**Corresponding author:**
[*]Stefan Schmid, PT, PhD, Bern University of Applied Sciences, Department of Health Professions, Murtenstrasse 10, 3008 Bern, Switzerland, +41 31 848 37 96, stefanschmid79@gmail.com

## ABSTRACT

Currently available musculoskeletal inverse-dynamics thoracolumbar spine models are entirely based on data from adults and might therefore not be applicable for simulations in children and adolescents. In addition, these models lack lower extremities, which are required for comprehensive evaluations of functional activities or therapeutic exercises. We therefore created OpenSim-based musculoskeletal full-body models including a detailed thoracolumbar spine for children and adolescents aged 6-18 years and validated by comparing model predictions to in vivo data.

After combining our recently developed adult thoracolumbar spine model with a lower extremity model, children and adolescent models were created for each year of age by adjusting segmental length and mass distribution, center of mass positions and moments of inertia of the major body segments as well as sagittal pelvis and spine alignment based on literature data. Similarly, muscle strength properties were adjusted based on CT-derived cross-sectional area measurements. Simulations were conducted from in vivo studies reported in the literature involving children and adolescents evaluating maximum trunk muscle strength (MTMS), lumbar disc compressibility (LDC), intradiscal pressure (IDP) and trunk muscle activity (MA).

Model predictions correlated highly with in vivo data (MTMS: $r \geq 0.82$, $p \leq 0.03$; LDC: $r=0.77$, $p<0.001$; IDP: $r \geq 0.78$, $p<0.001$; MA: $r \geq 0.90$, $p<0.001$), indicating suitability for the reasonably accurate prediction of maximal trunk muscle strength, segmental loading and trunk muscle activity in children and adolescents. When aiming at investigating children or adolescents with pathologies such as idiopathic scoliosis, our models can serve as a basis for the creation of deformed spine models and for comparative purposes.

# 1. INTRODUCTION

Musculoskeletal modeling of the spine allows predictions of segmental loading and individual muscle forces, which are usually not measurable *in vivo*. Gaining insight into such parameters might be crucial for a better understanding of spinal pathologies and to evaluate treatment effects. For example, vertebral compressive loads, as calculated by musculoskeletal models, may explain the higher incidence of fractures in thoracolumbar vertebrae compared to other spinal regions (Bruno et al., 2017a). However, currently available musculoskeletal inverse-dynamics thoracolumbar spine models (Bruno et al., 2015; Ignasiak et al., 2016) are entirely based on data from adults and might therefore not be applicable for simulations in children and adolescents. Especially when considering the non-linear development of anthropometric properties and spinal alignment during growth (Cil et al., 2005; Fryar et al., 2012; Jensen, 1989), the typical uniform scaling approach seems not appropriate to create such models. Moreover, the currently available thoracolumbar spine models are not full-body models, lacking lower extremities, which are required for comprehensive evaluations of functional activities or therapeutic exercises, especially in disorders where the spine is affected as a consequence of restrictions in the lower extremities or vice versa (Bangerter et al., 2018; Schmid et al., 2016a). Conversely, none of the existing validated full-body models (Actis et al., 2018; Bassani et al., 2017; Beaucage-Gauvreau et al., 2019; Raabe and Chaudhari, 2016) includes a fully articulated thoracolumbar spine to allow for detailed evaluation of spinal loading, and all are based on anthropometric properties derived from adults.

In order to conduct motion capture-based experiments aiming at the predictions of complex biomechanical parameters of the spine in children and adolescents, including those with musculoskeletal and neuromuscular disorders, validated full-body models including a detailed thoracolumbar spine are needed. Enhanced full-body joint kinematics such as previously obtained for different adolescent patient populations during walking (Bangerter et al., 2018; Schmid et al., 2016a; 2016b) could be used with such models to address advanced questions regarding the pathomechanisms of developmental disorders, ultimately leading to better preventive and treatment strategies. Therefore, the objectives of this work were to create male and female musculoskeletal full-body models including a detailed thoracolumbar spine and a rib cage for children and adolescents aged 6-18 years, to validate these models for predictions of trunk muscle strength, spinal loading and trunk muscle activity, and to make these models available for widespread use.

# 2. METHODS

## 2.1. Base models

Male and female versions of our recently developed and well-validated musculoskeletal model of a fully articulated adult thoracolumbar spine with a rib cage, a lumped head and neck body as well as upper extremities (Bruno et al., 2015; Bruno et al., 2017a) were used as the upper body base models. The models were created using the OpenSim musculoskeletal modeling environment (Delp et al., 2007) and contain over 30 muscle groups (552 individual muscle fascicles).

This upper body model was combined with the lower limbs of the Gait2354 model available in OpenSim (Anderson and Pandy, 1999, 2001; Carhart, 2000; Delp et al., 1990; Yamaguchi and Zajac, 1989). Mass and length properties of the legs were thereby scaled using the same anthropometric data that was used to create our thoracolumbar models (de Leva, 1996). Pelvis and sacrum center of mass (CoM) locations were retained from the Gait2354 model. Lower body inertial properties were manually scaled according to the segments' corresponding mass changes. Inertial properties for the thoracolumbar vertebrae and arms were added from literature (de Leva, 1996; Pearsall et al., 1996), after scaling for height and weight of our male and female models. The psoas muscle attachment points were modified from the thoracolumbar model so that the 4 most inferior points attached to the pelvis and femur as identified in the Gait 2354 model. Any other muscles connecting the pelvis to a body located inferiorly were incorporated from the Gait 2354 model, and any muscle connecting the pelvis to a body located superiorly were incorporated from the thoracolumbar spine model.

## 2.2. Children and adolescent models

The base (adult) full-body models were adjusted for each year of age between 6 and 18 years, resulting in a total of 26 children and adolescent models (13 male and 13 female). Table 1 provides an overview of the scaling information, whereas a complete set can be found in the Electronic Supplementary Material (Tables A1-5). Passive stiffness properties of the intervertebral joints were not modeled, since no appropriate data was available from the literature.

### *2.2.1. Anthropometric properties and sagittal alignment*

Body mass, body height and segmental length distribution for all models were scaled based on normative values obtained from the literature (Bundak et al., 2014; Fredriks et al., 2005; Fryar et al., 2012; Gleiss et al., 2013; Krogman, 1970; Zhang and Li, 2015; Zhu et al., 2015). For the male models segmental mass distribution, CoM positions, and moments of inertia of the major body segments were adjusted for each year of age using the regression equations provided by Jensen (1989). Unfortunately, this information was only available for boys, but the same equations were used to scale female segmental mass distribution, which has been shown to be similar between boys and girls (Yokoi et al., 1986). Due to lacking information, however, CoM positions and moments of inertia were not additionally adjusted for age in the female models.

In order to account for age-related changes in the sagittal alignment of the spine, sacrum orientation and intersegmental angles of the thoracolumbar spine were adjusted according to the available literature (Cil et al., 2005; Ghandhari et al., 2013; Kuntz et al., 2008; Mac-Thiong et al., 2004; Vedantam et al., 1998). Since no sex-related differences were found for sagittal alignment (Cil et al., 2005), the same values were used for boys and girls.

**Table 1:** Selection of parameters used to scale the children and adolescent models.

| Age [years] | Height [m] | | Upper body height / Height | | Mass [kg] | | Head+Neck mass / Mass | Upper trunk mass[1] / Mass | Lower trunk mass[2] / Mass | Pelvic incidence [°] | Lumbar lordosis [°] | Thoracic kyphosis [°] |
|---|---|---|---|---|---|---|---|---|---|---|---|---|
| | Boys | Girls | Boys | Girls | Boys | Girls | | | | | | |
| 6 | 1.18 | 1.17 | 0.48 | 0.48 | 22.2 | 21.5 | 0.17 | 0.15 | 0.27 | 43.9 | 50.6 | 37.2 |
| 7 | 1.24 | 1.23 | 0.48 | 0.47 | 23.9 | 24.1 | 0.16 | 0.15 | 0.27 | 44.4 | 54.6 | 38.2 |
| 8 | 1.29 | 1.28 | 0.47 | 0.47 | 27.7 | 27.5 | 0.14 | 0.14 | 0.27 | 45.0 | 54.6 | 38.2 |
| 9 | 1.35 | 1.34 | 0.47 | 0.46 | 30.9 | 29.8 | 0.13 | 0.14 | 0.27 | 45.5 | 54.6 | 38.2 |
| 10 | 1.40 | 1.40 | 0.46 | 0.46 | 35.3 | 35.4 | 0.12 | 0.14 | 0.27 | 47.0 | 60.8 | 40.2 |
| 11 | 1.45 | 1.47 | 0.46 | 0.46 | 38.9 | 39.3 | 0.11 | 0.14 | 0.27 | 47.7 | 60.8 | 40.2 |
| 12 | 1.51 | 1.53 | 0.45 | 0.45 | 43.4 | 44.6 | 0.10 | 0.14 | 0.27 | 48.5 | 60.8 | 40.2 |
| 13 | 1.59 | 1.57 | 0.45 | 0.45 | 50.9 | 48.7 | 0.09 | 0.14 | 0.27 | 49.2 | 59.9 | 43.3 |
| 14 | 1.65 | 1.60 | 0.45 | 0.46 | 56.4 | 53.7 | 0.09 | 0.15 | 0.27 | 50.0 | 59.9 | 43.3 |
| 15 | 1.70 | 1.62 | 0.45 | 0.46 | 61.7 | 57.0 | 0.08 | 0.15 | 0.27 | 50.7 | 59.9 | 43.3 |
| 16 | 1.73 | 1.63 | 0.45 | 0.46 | 65.6 | 57.5 | 0.08 | 0.16 | 0.27 | 51.5 | 59.9 | 43.3 |
| 17 | 1.75 | 1.64 | 0.46 | 0.46 | 68.1 | 58.9 | 0.07 | 0.16 | 0.27 | 52.2 | 59.9 | 43.3 |
| 18 | 1.76 | 1.63 | 0.46 | 0.46 | 75.2 | 56.8 | 0.07 | 0.17 | 0.27 | 52.9 | 59.9 | 43.3 |

[1]Without head, neck and arms.

[2]Including the pelvis.

### 2.2.2. Muscle strength properties

The morphology of the erector spinae (ES), multifidi (MF), psoas major (PM) and quadratus lumborum (QL) muscles was adjusted using CT-based L3-S1 cross-sectional area (CSA) measurements from healthy children and adolescents aged 5-20 years (Been et al., 2018). Since these values were only available as averages of both gender, percentage differences from weighted averages (using male-female ratio of the reported children and adolescent CSAs) of the male and female adult model CSAs were used to estimate CSAs for boys and girls separately. For the trunk muscles without measured CSA data, scaling factors were established based on the available data. Muscle strength in the lower and upper extremities as well as the neck was scaled using a general scale factor, which consisted of the average value of the factors used to scale ES, MF and QL at L3-4 as well as PM at L3-5.

Adjusted trunk muscle CSAs were subsequently "fine-tuned" by comparing simulations of maximal trunk muscle strength (MTMS) to *in vivo* data from the literature (Andersen and Henckel, 1987; Peltonen et al., 1998; Sinaki et al., 1996). Details of the MTMS simulations are described in section 2.3.1., whereas more information on the CSA scaling can be found in Bruno et al. (2015). To convert trunk muscle CSA into fascicle strength, a uniform maximal muscle stress (MMS) of 100 N/cm$^2$ was used (Ballak et al., 2014; Bruno et al., 2015; Bruno et al., 2017b; Burkhart et al., 2018; O'Brien et al., 2010).

## 2.3. Validation studies

Following best practice guidelines for verification and validation of musculoskeletal models (Hicks et al., 2015), we sought to ensure appropriate musculoskeletal geometry, trunk strength, spinal loading, and muscle activation patterns in comparison to measured values. Thus, simulations were conducted from published *in vivo* studies using OpenSim 3.3 (Delp et al., 2007) and MATLAB R2017a (MathWorks Inc., Natick, MA, USA). For each validation study, scaled models were created corresponding to gender, age, height and weight of the subjects examined in the *in vivo* studies. Joint angles were derived from the information provided in the respective *in vivo* studies (details in Tables A6-8 in the Electronic Supplementary Material). All models were solved using an inverse dynamics based static optimization with a cost function that minimized the sum of squared muscle activation (Herzog, 1987). A weighted average from the values predicted by the male and female models was calculated based on the male-female ratios provided in the respective studies, except for maximal trunk muscle strength where male and female values were reported separately. Validity was investigated using Pearson's correlations and simple linear regressions. Statistical significance was accepted at an alpha level of 0.05.

### 2.3.1. Maximal trunk muscle strength (MTMS)

To determine the MTMS of our models, *in vivo* data from 6-18 years old children and adolescents conducting maximum voluntary isometric contractions against a force transducer in upright standing (Andersen and Henckel, 1987; Peltonen et al., 1998) as well as prone and supine positions (Sinaki et al., 1996) were used. For validation of prone and supine MTMS, in vivo values were established using the provided regression equations with our models' gender, height and mass. After placing the models in the respective positions, increasing external forces (10 N increments) were applied to the spine (standing: T5-6; prone: T1-3) to evaluate extensor strength and to the sternum (standing: top of lower third; supine: base) to evaluate flexor strength (Figs. 1-2). To simulate the floor and the wooden plate used to stabilize the pelvis in standing as well as the examining table in prone and supine positions, we applied residual point actuators with a maximal force of 10 kN, which represents a large enough force to provide the required support with minimal expenses in the static optimization. Once the static optimization started to fail consistently, the models were considered too weak to resist the applied forces and thus, MTMS was defined as the highest applied force for which the static optimization successfully solved. This approach was previously used by our group to determine MMS of trunk extensors in older adults (Burkhart et al., 2018).

*2.3.2. Lumbar disc compressibility (LDC)*

MRI-based *in vivo* measurements of LDC (or lumbar disc height change) due to 4, 8 and 12 kg backpack loads in a group of adolescents with an average age of 11 years (Neuschwander et al., 2010) were used to evaluate the accuracy of the compressive forces predicted by our models. For each loading condition, we bilaterally applied a 20° dorsally angled external force (16.6, 39.3 and 58.9 N, respectively) to the lateral third of the upper edge of the scapula and calculated the compressive forces on the levels T12/L1 to L5/S1. To convert the forces to disc height change, we used a lumbar disc force-displacement curve that was available from the literature (Markolf, 1972). LDC values of the loaded conditions were expressed as a percentage of unloaded upright standing.

*2.3.3. Intradiscal pressure (IDP)*

Lumbar (L3/4) and thoracic (middle: T6/7, T7/8; lower: T9/10, T10/11) IDPs were evaluated by placing an 18 years old male and female model in different standing positions and by applying external forces (Table 2) as conducted *in vivo* with a pressure-sensing needle that was inserted into the nucleus pulposa of the respective lumbar disc in 19-23 year olds (Schultz et al., 1982) and thoracic discs in 19-47 year olds (Polga et al., 2004). We selected these two studies because no *in vivo* studies were available involving individuals aged 6-18 years and these studies involved individuals closest to this age range. To compare the predicted vertebral compressive forces to in vivo measured IDPs, compressive forces ($F_{Comp}$) were converted into IDPs using the formula

$$IDP = \frac{F_{Comp}}{CSA_{Vert} \times 0.66} \tag{1}$$

where $CSA_{Vert}$ is the CSA of the vertebral body endplate and 0.66 a correction factor for translating between IDP and compressive force as previously described (Bruno et al., 2015; Dreischarf et al., 2013; Nachemson, 1960, 1966). We thereby used vertebral body endplate CSAs of 17.4 $cm^2$ for L4 (Been et al., 2018) as well as 9.6/7.7 $cm^2$ (males/females) for T10-11 and 7.1/5.7 $cm^2$ for T7-8 (Kishimoto et al., 2016).

*2.3.4. Trunk muscle activity (TMA)*

The accuracy of TMA predictions was investigated using surface electromyography (EMG) data that was measured along with the above described lumbar IDP (Schultz et al., 1982). Predictions of erector spinae, rectus abdominis and abdominal oblique muscle activation were therefore derived from the same simulations as conducted for the evaluation of lumbar IDP (Table 2). Muscle fascicles of the model were selected according to the placement of the surface electrodes in the study. TMA for each muscle was expressed as a percentage of unloaded upright standing and presented as an average of the left and right sides.

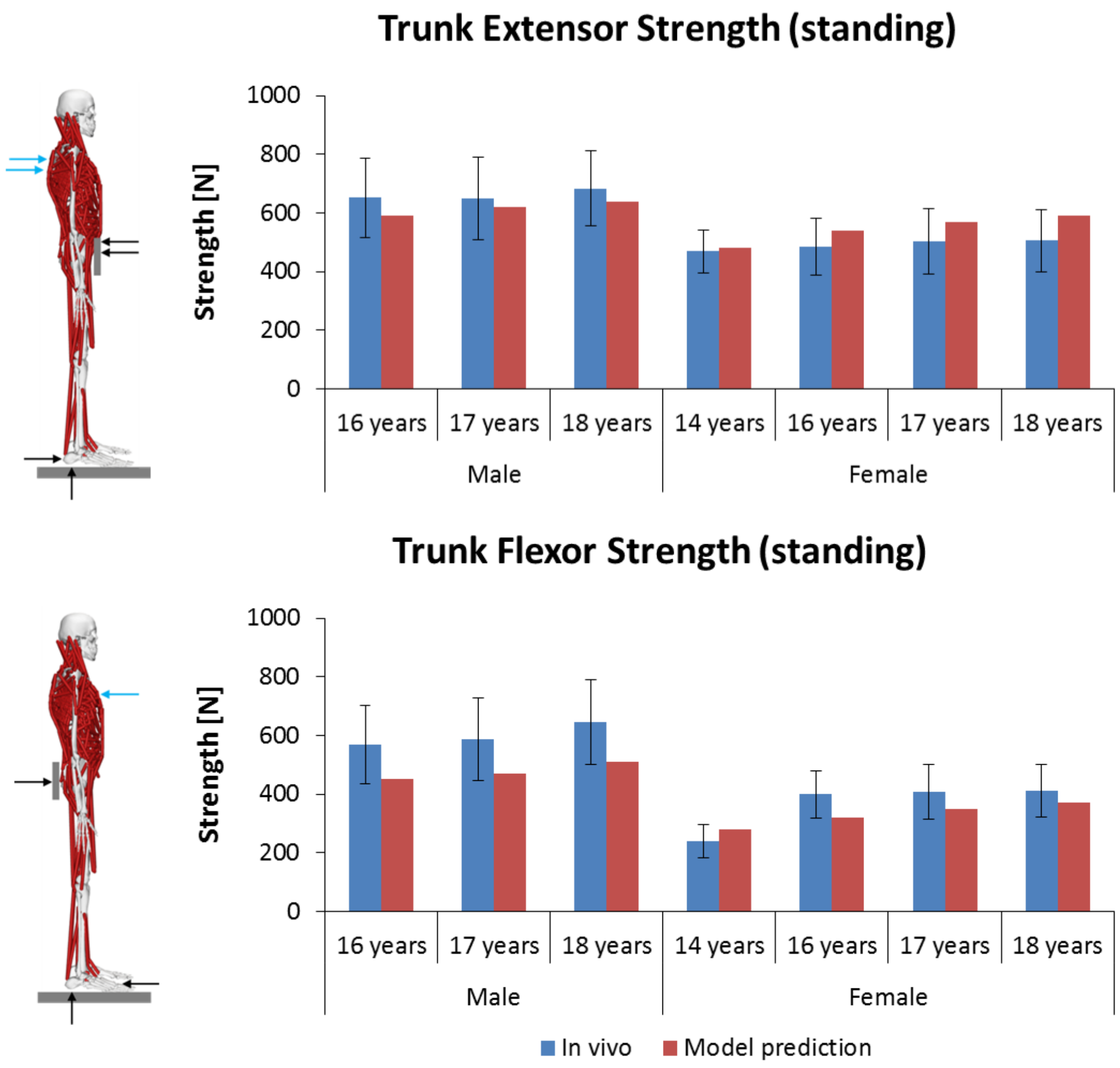

**Figure 1:** Trunk extensor and flexor strength predictions vs. in vivo data for 16-18 year old males and 14-18 year old females in a standing position. In vivo data was retrieved from Peltonen et al. (1998) for the 14 year old females and from Andersen et al. (1987) for the 16-18 years old males. The error bars indicate 1 SD above and below the mean of the in vivo data. The light blue arrows pointing on the model on the left indicate the external forces that were applied on the spine (top) and the sternum (bottom), whereas the black arrows indicate the residual point actuators used to simulate the floor and the wooden plate that stabilized the pelvis in the in vivo measurements.

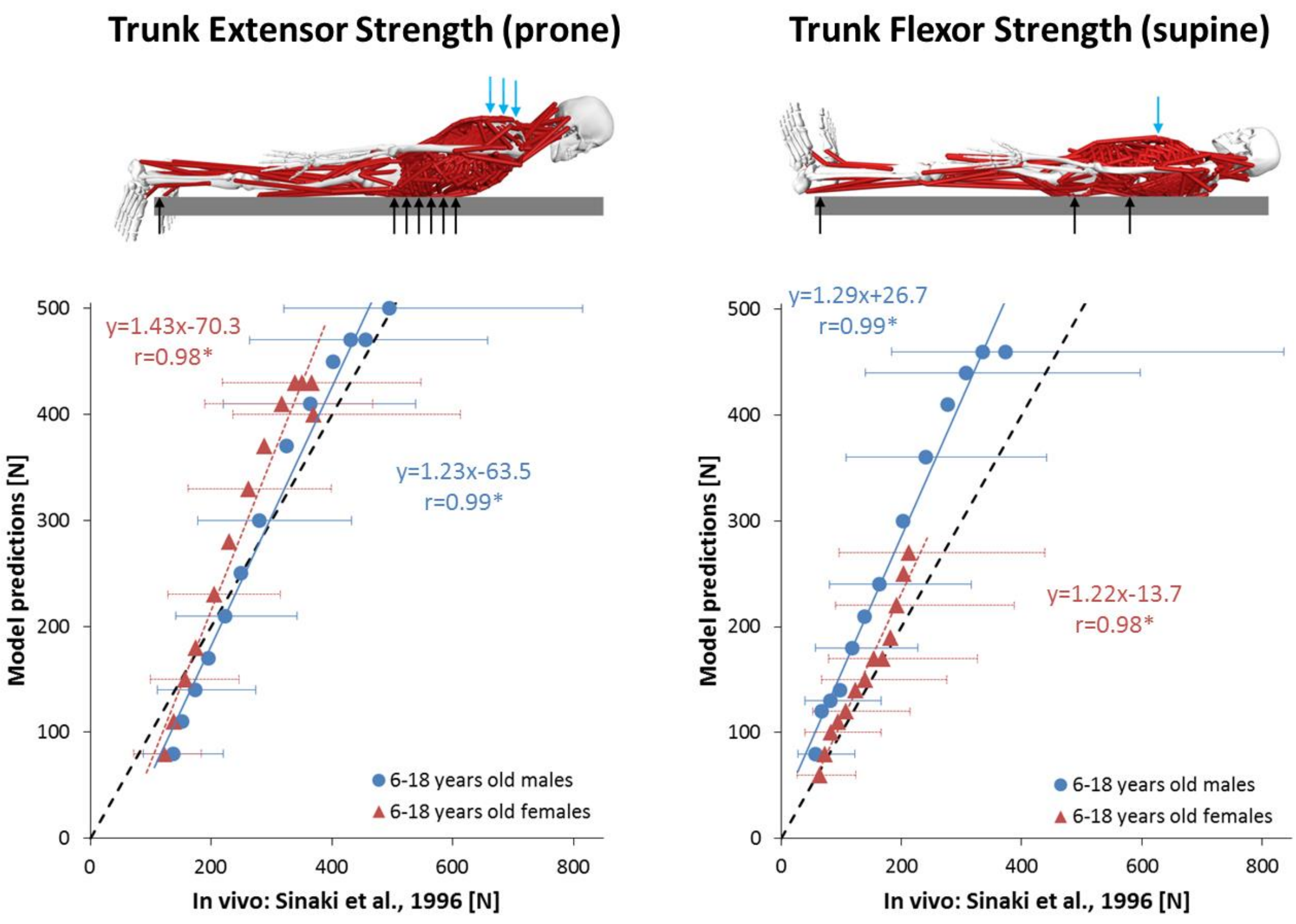

**Figure 2:** Trunk extensor and flexor strength predictions vs. in vivo data for 6-18 year old males and females in a prone and supine position. For the in vivo data, muscle strength values were calculated based on the regression equations provided by Sinaki et al. (1996) using our models' gender, height and mass. The error bars indicate the 95% normative limits of the 246 subjects that were actually measured in vivo. The light blue arrows pointing on the model on the top indicate the external forces that were applied on the spine (left) and the sternum (right), whereas the black arrows indicate the residual point actuators used to simulate the examining table in the in vivo measurements. The dashed black lines represent unity. The asterisks (*) indicate statistically significant correlations ($p<0.001$).

**Table 2:** Simulations conducted for the prediction of intradiscal pressure (IDP).

| | |
|---|---|
| **Lumbar IDP** | Relaxed upright standing. |
| (from: Schultz et al., 1982) | Resisting a flexion force of 147.2 N (15 kg) applied dorsally on the vertebrae T5-6. |
| | Resisting a right lateral-bending force of 147.2 N (15 kg) applied laterally on the thorax at the height of the 5$^{th}$ rib. |
| | Resisting an extension force of 147.2 N (15 kg) applied ventrally on the lower third of the sternum. |
| | Resisting a right twisting force of two times 98.1 N (10 kg) applied in opposite directions on both sides of the thorax at the height of the 5$^{th}$ rib. |
| | Upright standing with the hands close to the chest and holding an 8kg weight equally distributed between the hands (2 x 39.2 N). |
| | Upright standing with the arms extended forward. |
| | Upright standing with the arms extended forward and holding an 8 kg weight equally distributed between the hands (2 x 39.2 N). |
| | Standing with both hips flexed 30° and with the arms extended forward. |
| | Standing with both hips flexed 30° and with the arms extended forward and holding an 8 kg weight equally distributed between the hands (2 x 39.2 N). |
| **Thoracic IDP** | Relaxed upright standing. |
| (from: Polga et al., 2004) | Standing with the lumbar spine 15° extended. |
| | Standing with the lumbar spine 20° laterally bended to the right. |
| | Standing with 30° trunk forward flexion. |
| | Standing with the thoracolumbar spine 30° axially rotated to the left side. |
| | Standing with the arms in neutral position and holding a 10 kg weight (98.1 N) in each hand. |
| | Standing with the elbows 90° flexed and holding a 10 kg weight (98.1 N) in each hand. |
| | Standing with 30° trunk forward flexion and holding a 10 kg weight (98.1 N) in each hand (arms hanging down vertically). |

## 3. RESULTS

### *3.1. Muscle strength properties*

The adjustment of muscle strength properties based on trunk muscle CSA data and MTMS simulations resulted in age- and gender-specific maximum strength values for each muscle fascicle in the models. Fig. 3 illustrates the final CSAs for the erector spinae muscle in boys and girls.

### *3.2. Validation studies*

Predicted MTMS in a standing position was correlated highly (extensors: r=0.82, p=0.03; flexors: r=0.98, p<0.001) with and within 1 SD of the *in vivo* measurements for 16-18 years old boys and 14-18 years old girls (Fig. 1). In prone and supine positions, MTMS for 6-18 years old boys and girls showed very high correlations (r≥0.98, p<0.001) between model predictions and *in vivo* measurements, with the majority of estimated strength values being well within the 95% normative limits reported for the 246 experimentally evaluated subjects (Fig. 2). Slope and intercept of the fitted regression line for male trunk flexor strength indicated constant overestimation of the model predictions. For all other strength parameters, regression line characteristics indicated particularly high MTMS prediction accuracy around the ages 8-10 years and a tendency for overestimation with increasing and underestimation with decreasing ages.

Simulations of LDC for different backpack loads correlated as well highly with *in vivo* data ($r=0.77$, $p<0.001$) (Fig. 4). Predictions for the levels T12/L1, L1/2 and L3/4 showed high accuracy (largely within 1 SD of in vivo measurements), whereas for L4/5 and L5/S1, LDC tended to be under predicted. LDC at L2/3 showed high accuracy for a 4kg backpack load but tended to be over predicted for 8kg and 12kg.

For IDP, model predictions correlated highly ($r \geq 0.78$, $p<0.001$) with *in vivo* measurements for the lumbar as well as thoracic regions (Fig. 5). However, slope and intercept of the fitted regression line for lumbar IDP indicated overestimation for most activities.

The evaluation of TMA predictions showed high correlations ($r \geq 0.90$, $p<0.001$) with *in vivo* measurements for the erector spinae and abdominal oblique muscles (Fig. 6). Rectus abdominis activation could not be evaluated, since the model predicted an activation level of almost zero for all tasks except from resisting lateral-bending, extension and twisting forces.

Individual IDPs and TMAs corresponding to the simulated activities can be found in the Electronic Supplementary Material (Figs. A1-A5).

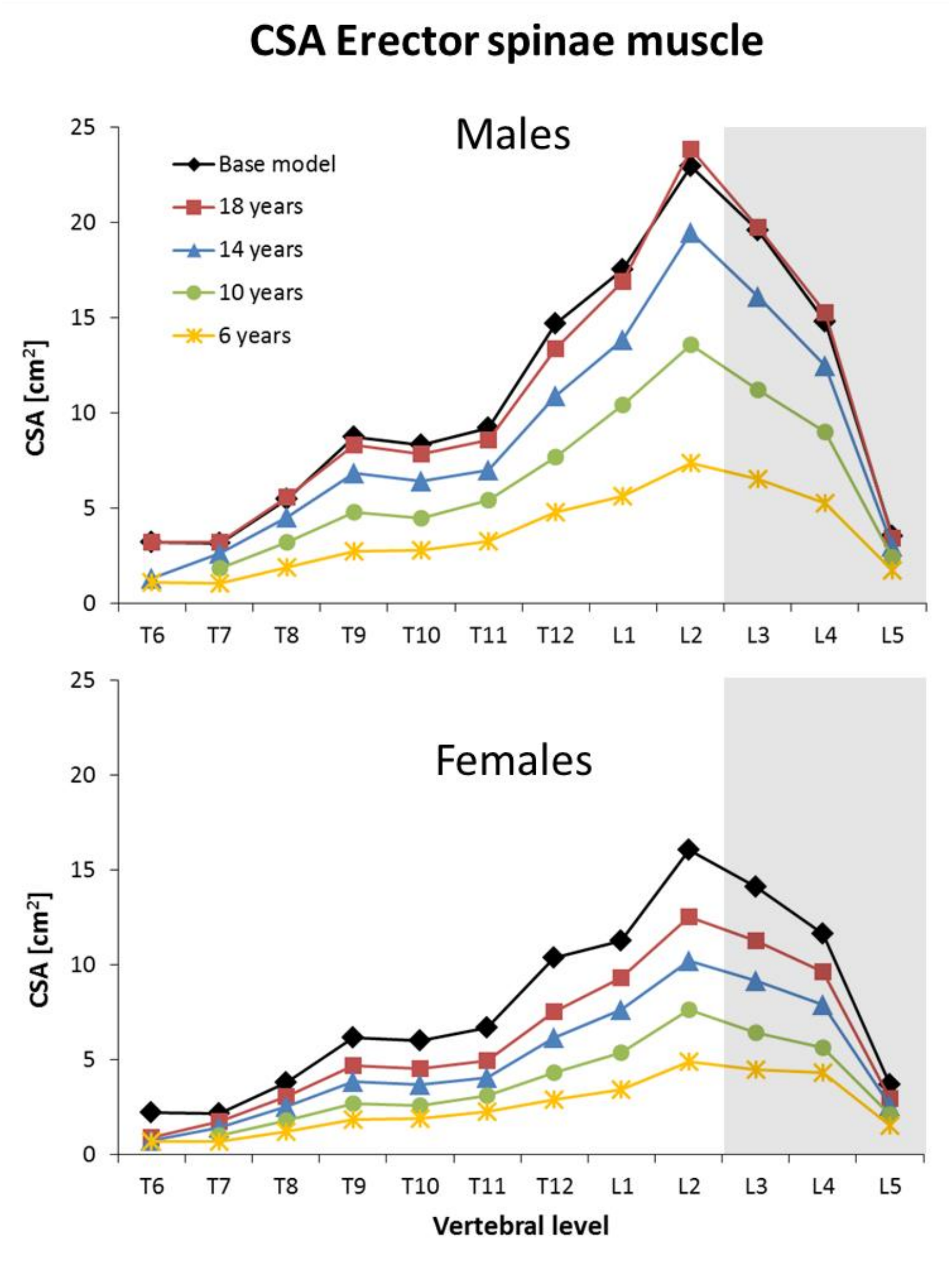

**Figure 3:** Exemplary illustration of cross-sectional areas (CSA) of the erector spinae muscle in the male and female models for the ages 6, 10, 14 and 18 years as well as for the base models. The gray shaded areas indicate the vertebral levels that were scaled directly based on data from the literature (Been et al., 2018).

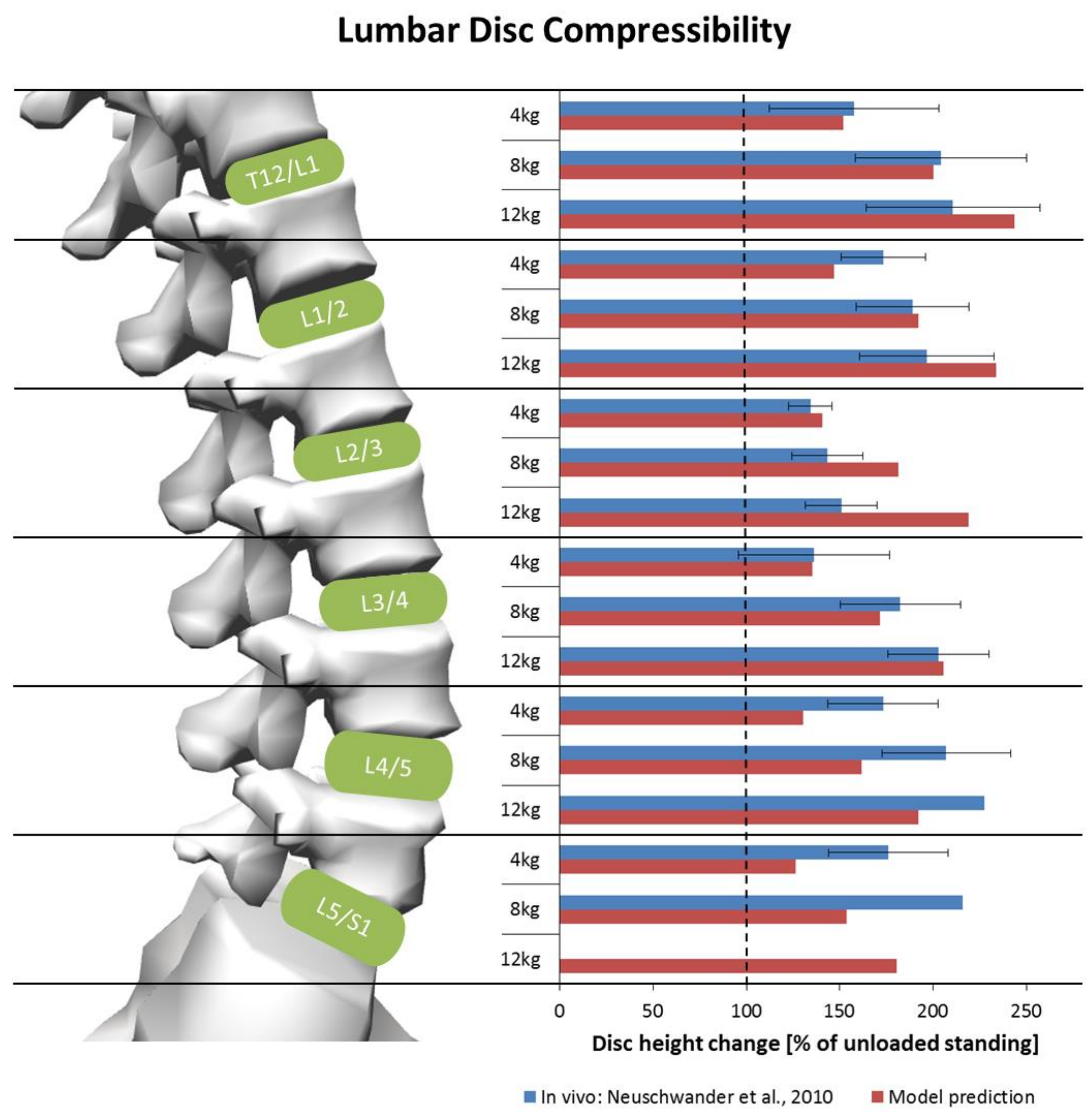

**Figure 4:** Model predictions vs. in vivo data of lumbar disc compressibility at the T12/L1, L1/2, L2/3, L3/4, L4/5 and L5/S1 levels for 11 year old males and females carrying 4, 8 and 12 kg backpack loads while standing in an upright MRI scanner (Neuschwander et al., 2010). All values are expressed as a percentage of unloaded upright standing, which is represented by the dotted vertical line. The error bars indicate 1 SD above and below the mean of the in vivo data.

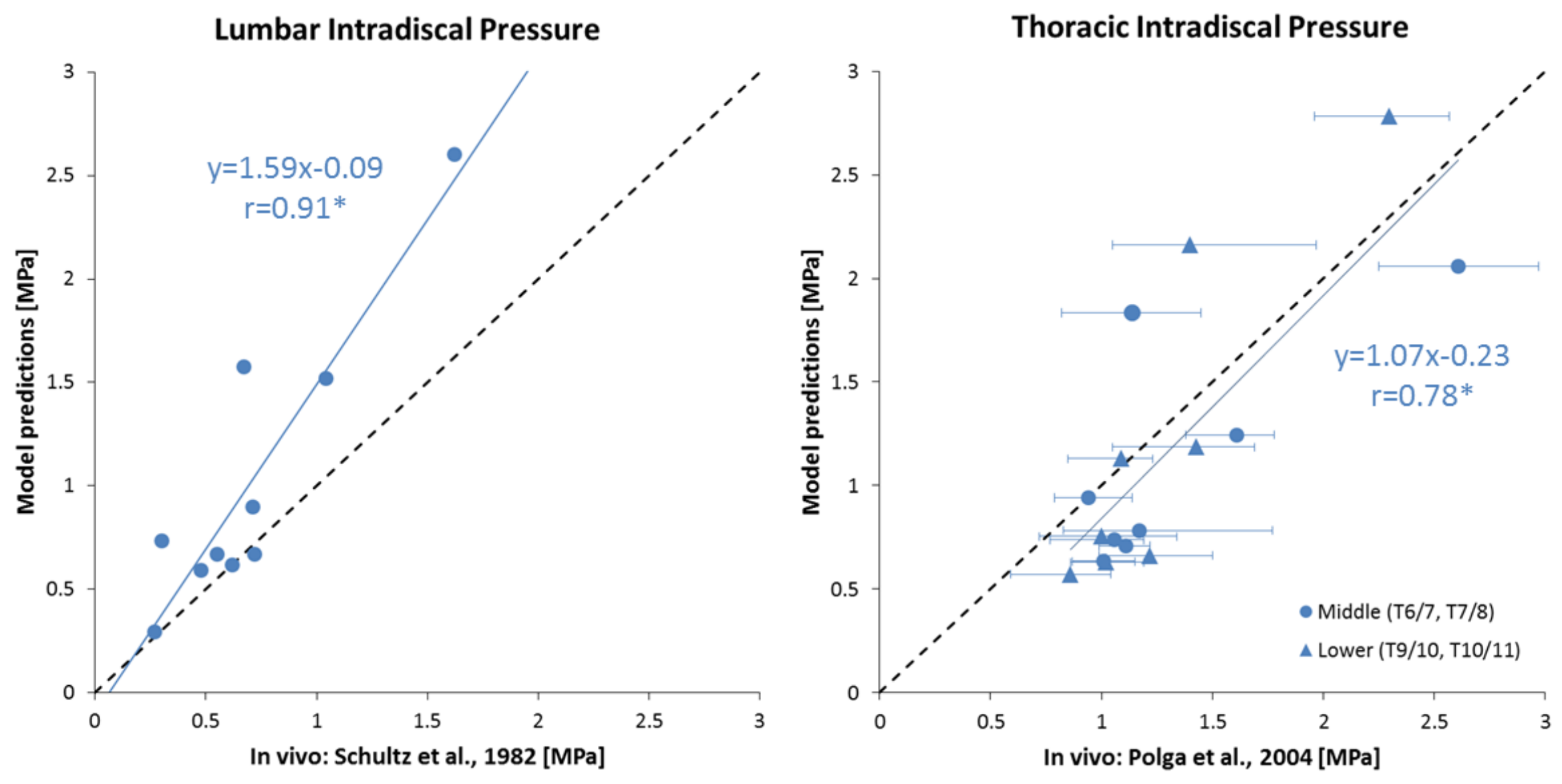

**Figure 5:** Model predictions of intradiscal pressure vs. in vivo data obtained using a pressure-sensing needle inserted into the L3/4 disc in 19-23 year old (Schultz et al., 1982) and into several thoracic levels in 19-47 year old males and females (Polga et al., 2004). The error bars indicate the range of the measurements. The dashed black lines represent unity. The asterisks (*) indicate statistically significant correlations ($p<0.001$).

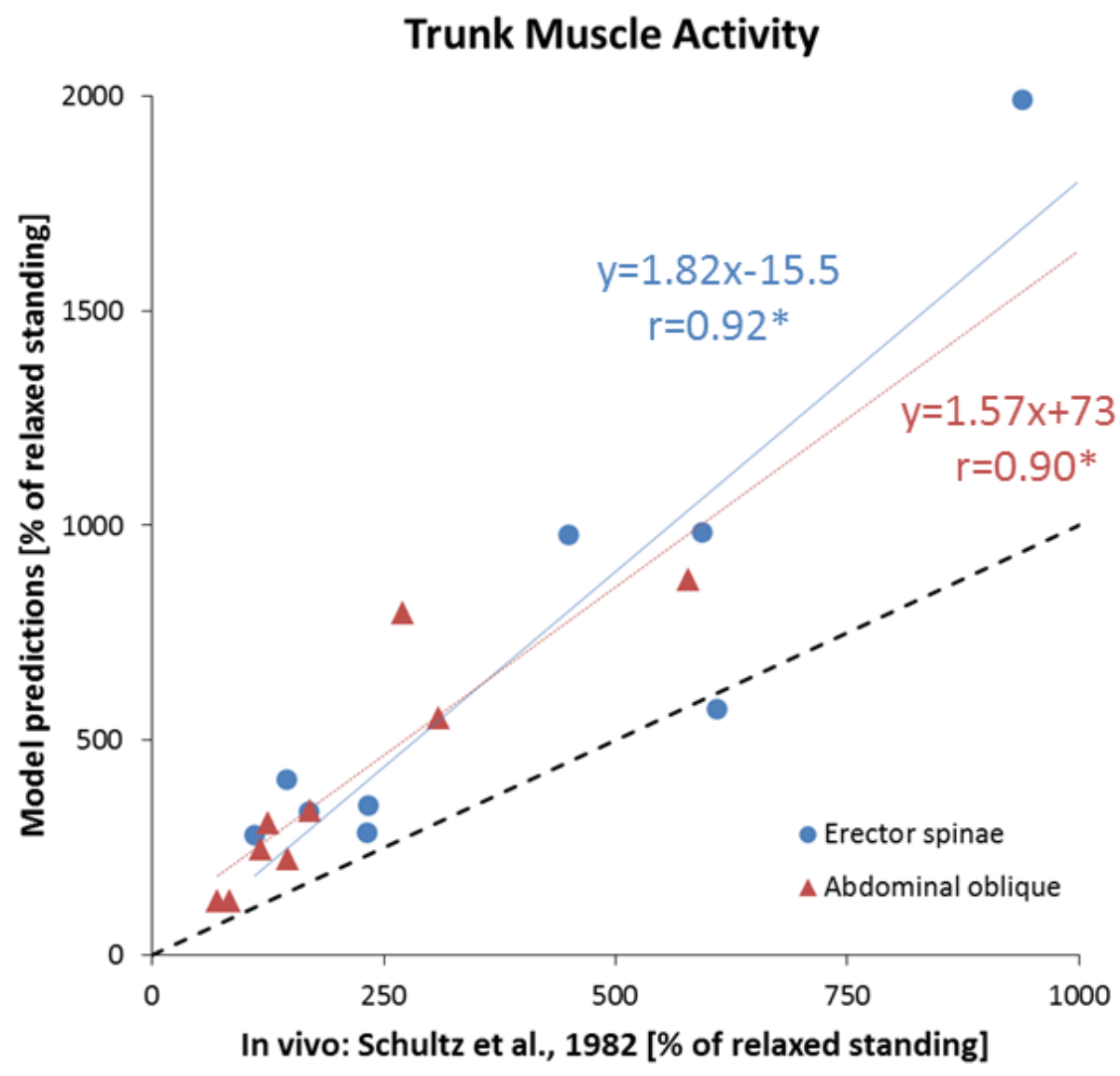

**Figure 6:** Erector spinae and abdominal oblique muscle activity predictions vs. in vivo measurements obtained from surface electromyography in a group of 19-23 year old males and females (Schultz et al., 1982). The dashed black lines represent unity. The asterisks (*) indicate statistically significant correlations ($p<0.001$).

## 4. DISCUSSION

We created and validated male and female musculoskeletal full-body models including a detailed thoracolumbar spine and a rib cage for children and adolescents aged 6-18 years. Since previous thoracolumbar spine models and full-body models were entirely based on data from adults these are the first of their kind.

Special features are the adjustment of muscle morphology and sagittal alignment including sacral slope for each year of age between 6-18 years. Previous work by our group showed that trunk muscle morphology and sagittal spinal shape has a major impact on estimated vertebral compressive and shear loads and should therefore always be accounted for (Bruno et al., 2017b). Of course, even higher prediction accuracy could be reached by scaling these properties based on subject-specific information. It has to be considered though that such information is not always readily available and could entail high costs, which is a limiting factor in clinical as well as research settings.

The high correlations of the models' predicted MTMS with *in vivo* measured maximum voluntary isometric contractions imply that the age-dependent adjustment of trunk muscle CSA together with sagittal alignment and segmental mass distribution was conducted in an appropriate manner.

The distinct overestimation of male flexor strength in supine position might be attributable to the assumption of a uniform MMS of 100 N/cm$^2$ for all trunk muscles as well as the age-dependent adjustment of segmental mass distribution that was only available for males in the literature but applied in the same way for both gender in our models.

A recently conducted systematic review indicated that MMS was about 20% higher in 25 years old young adults when compared to 76 years old men (Ballak et al., 2014). Considering the facts that the average trunk extensor MMS for elderly adults (65 years and older) without back pain was reported to be 78.3 N/cm$^2$, and that no differences were reported between 28 years old adults and 9 years old children or between males and females (Burkhart et al., 2018; O'Brien et al., 2010), the general assumption of an MMS of 100 N/cm$^2$ seems appropriate. However, the findings presented by Ghezelbash et al. (2018) suggested that MMS values might be lower for certain trunk flexor (rectus abdominis: 40±22 N/cm$^2$; iliopsoas: 69±42 N/cm$^2$) than trunk extensor muscles (longissimus: 92±42 N/cm$^2$; multifidi: 81±38 N/cm$^2$), indicating a clear need for future research targeting the more detailed definition of individual trunk muscle MMS in all age groups.

Predicted LDC values correlated well with *in vivo* measurements, highlighting the potential of our models for the estimation of parameters beyond compressive forces. The respective over- and underestimation of LDC at L2/3 (8 and 12kg), L4/5 and L5/S1, might most likely be linked to the conversion of compressive forces into disc height change using a force-displacement curve that was derived from 26 isolated adult cadaveric discs (ages 21-55 years, levels T12/L1 to L3/4) (Markolf, 1972). It is likely that the mechanical properties of lumbar intervertebral discs differ between adults and 11 year old adolescents as well as among the different spinal levels, which is why these results should be treated with great caution.

The analysis of estimated IDP yielded high correlations with *in vivo* measurements, similar as previously reported for the standalone adult thoracolumbar spine model (Bruno et al., 2015). The tendency for over-prediction of lumbar IDP might be explained by the fact that the intervertebral joints were modeled as simple ball joints located at the geometric center of the intervertebral disc and no facet joints were included. Previous research showed that lumbar facet joints carry up to 25% of the compressive load applied to the spine in a neutral posture, with load contribution increasing during extension and decreasing during flexion (Jaumard et al., 2011).

Another possible explanation for the overestimation of lumbar IDP could be the omission of intraabdominal pressure (IAP) in our models. It was shown that during extension efforts, IAP associated with abdominal muscle activation produced spinal unloading by generating an extension moment that exceeded the flexion moment of the abdominal wall muscle activation forces (Stokes et al., 2010).

The fact that our simulations were conducted based on joint angles derived from the literature rather than motion capture data might as well have compromised the prediction accuracy of lumbar IDP. Bassani et al., (2017), for example, reported slightly more accurate predictions of lumbar IDP (similar correlations coefficients, but closer to unity) using a motion capture-driven full-body model. It has to be considered, however that their validation involved a different lumbar level (L4/5) as well as a different set of functional activities.

Furthermore, the omission of IAP in our models might (at least partially) explain the inaccurate predictions for rectus abdominis muscle activation. Without IAP, there was no need for our models to increase rectus abdominis muscle activation during activities that did not involve a "direct provocation" of the rectus abdominis muscle (i.e. resisting an extension force). Predicted erector spinae and abdominal oblique muscles activity, on the other hand, proved to be suitable for capturing the relative changes among the different activities.

Major limitations in the creation of our children and adolescent models include the facts that CSA data were only available for 4 trunk muscles, only for the levels L3-S1 and only as average values between boys and girls. In addition, the fact that the literature provided no specific information on inertial properties in girls may limit the accuracy for dynamic simulations of functional activities in this population. It is worth mentioning though that the scale factors found for boys based on the literature were very close to 1, and thus did not result in substantially different inertial properties when applied. Moreover, the fact that the current models do not contain intervertebral stiffness properties might limit their use for simulations of dynamic activities.

In conclusion, we created musculoskeletal full-body models including a detailed thoracolumbar spine for boys and girls aged 6-18 years, which are freely available on the project-hosting platform SimTK (https://simtk.org/projects/spine-children). Despite certain limitations, the validation studies indicate suitability of these models for the reasonably accurate prediction of maximal trunk muscle strength, segmental loading and trunk muscle activity in children and adolescents. When aiming at investigating children or adolescents with pathologies such as idiopathic scoliosis, our models can serve as a basis for the creation of deformed spine models as well as for comparative purposes. Further development should include the implementation of IAP and intervertebral joint stiffness, especially when aiming at simulations of dynamic activities.

## 5. CONFLICT OF INTEREST STATEMENT

The authors declare no conflicts of interest.

## 6. ACKNOWLEDGMENTS

This work was funded by the Swiss National Science Foundation (SNSF, grant-no.: 178427) and the National Center for Simulation in Rehabilitation Research (NCSRR, sub-award of NIH grant no. 5P2CHD065690).

**Table A1:** Orange: Total body height [cm]. Green: Length proportions for the major body segments.

| Sex/Age | Height | Segmental length proportions | | | | | |
|---|---|---|---|---|---|---|---|
| | | Trunk+Head+Neck | Thigh | Shank | UpperArm | Forearm | Hand |
| **Male** | | | | | | | |
| 6 | 118.2 | 0.484 | 0.234 | 0.233 | 0.183 | 0.141 | 0.113 |
| 7 | 123.8 | 0.478 | 0.235 | 0.238 | 0.183 | 0.141 | 0.114 |
| 8 | 129.2 | 0.472 | 0.233 | 0.243 | 0.183 | 0.143 | 0.114 |
| 9 | 134.5 | 0.469 | 0.233 | 0.247 | 0.186 | 0.144 | 0.111 |
| 10 | 140.1 | 0.464 | 0.235 | 0.250 | 0.187 | 0.144 | 0.113 |
| 11 | 145.4 | 0.457 | 0.236 | 0.255 | 0.188 | 0.144 | 0.112 |
| 12 | 150.9 | 0.453 | 0.242 | 0.255 | 0.191 | 0.146 | 0.111 |
| 13 | 158.7 | 0.450 | 0.240 | 0.257 | 0.194 | 0.149 | 0.115 |
| 14 | 165.2 | 0.451 | 0.240 | 0.258 | 0.194 | 0.146 | 0.112 |
| 15 | 170.2 | 0.453 | 0.241 | 0.256 | 0.196 | 0.146 | 0.113 |
| 16 | 172.6 | 0.455 | 0.240 | 0.256 | 0.199 | 0.147 | 0.113 |
| 17 | 174.6 | 0.457 | 0.238 | 0.257 | 0.197 | 0.148 | 0.113 |
| 18 | 175.5 | 0.458 | 0.239 | 0.256 | 0.200 | 0.149 | 0.113 |
| **Female** | | | | | | | |
| 6 | 117.4 | 0.480 | 0.220 | 0.248 | 0.182 | 0.139 | 0.112 |
| 7 | 123.0 | 0.472 | 0.223 | 0.254 | 0.182 | 0.140 | 0.114 |
| 8 | 128.3 | 0.467 | 0.227 | 0.254 | 0.183 | 0.140 | 0.113 |
| 9 | 134.0 | 0.462 | 0.230 | 0.256 | 0.186 | 0.144 | 0.112 |
| 10 | 140.3 | 0.458 | 0.233 | 0.257 | 0.187 | 0.142 | 0.112 |
| 11 | 146.6 | 0.455 | 0.238 | 0.257 | 0.188 | 0.141 | 0.110 |
| 12 | 152.6 | 0.454 | 0.235 | 0.261 | 0.192 | 0.144 | 0.110 |
| 13 | 157.0 | 0.454 | 0.238 | 0.260 | 0.195 | 0.144 | 0.111 |
| 14 | 160.3 | 0.455 | 0.238 | 0.257 | 0.193 | 0.145 | 0.111 |
| 15 | 161.9 | 0.457 | 0.237 | 0.257 | 0.193 | 0.144 | 0.112 |
| 16 | 162.7 | 0.459 | 0.239 | 0.254 | 0.195 | 0.145 | 0.112 |
| 17 | 163.6 | 0.460 | 0.236 | 0.255 | 0.194 | 0.143 | 0.112 |
| 18 | 163.4 | 0.462 | 0.234 | 0.256 | 0.196 | 0.144 | 0.112 |

**Table A2:** Orange: Total body mass [kg]. Green: Mass proportions for the major body segments.

| Sex/Age | Total | Segmental mass proportions | | | | | | | | |
|---|---|---|---|---|---|---|---|---|---|---|
| | | HeadNeck | UpTrunk | LowTrunkPelvis | UpLeg | LowLeg | Foot | UpperArm | Forearm | Hand |
| **Male** | | | | | | | | | | |
| 6 | 22.2 | 0.171 | 0.152 | 0.273 | 0.087 | 0.044 | 0.020 | 0.028 | 0.015 | 0.009 |
| 7 | 23.9 | 0.157 | 0.147 | 0.273 | 0.092 | 0.046 | 0.020 | 0.028 | 0.016 | 0.009 |
| 8 | 27.7 | 0.144 | 0.144 | 0.273 | 0.097 | 0.048 | 0.021 | 0.029 | 0.016 | 0.009 |
| 9 | 30.9 | 0.132 | 0.142 | 0.273 | 0.101 | 0.050 | 0.021 | 0.030 | 0.016 | 0.009 |
| 10 | 35.3 | 0.120 | 0.141 | 0.273 | 0.105 | 0.051 | 0.021 | 0.030 | 0.017 | 0.009 |
| 11 | 38.9 | 0.110 | 0.141 | 0.273 | 0.108 | 0.052 | 0.021 | 0.031 | 0.017 | 0.009 |
| 12 | 43.4 | 0.102 | 0.142 | 0.273 | 0.110 | 0.053 | 0.021 | 0.032 | 0.017 | 0.009 |
| 13 | 50.9 | 0.094 | 0.144 | 0.273 | 0.112 | 0.053 | 0.020 | 0.032 | 0.017 | 0.009 |
| 14 | 56.4 | 0.087 | 0.148 | 0.273 | 0.114 | 0.052 | 0.020 | 0.033 | 0.018 | 0.009 |
| 15 | 61.7 | 0.081 | 0.152 | 0.273 | 0.115 | 0.052 | 0.019 | 0.034 | 0.018 | 0.009 |
| 16 | 65.6 | 0.076 | 0.158 | 0.273 | 0.115 | 0.051 | 0.019 | 0.035 | 0.018 | 0.009 |
| 17 | 68.1 | 0.073 | 0.165 | 0.273 | 0.115 | 0.049 | 0.018 | 0.035 | 0.019 | 0.009 |
| 18 | 75.2 | 0.070 | 0.173 | 0.273 | 0.114 | 0.048 | 0.017 | 0.036 | 0.019 | 0.009 |
| **Female** | | | | | | | | | | |
| 6 | 21.5 | 0.171 | 0.152 | 0.273 | 0.087 | 0.044 | 0.020 | 0.028 | 0.015 | 0.009 |
| 7 | 24.1 | 0.157 | 0.147 | 0.273 | 0.092 | 0.046 | 0.020 | 0.028 | 0.016 | 0.009 |
| 8 | 27.5 | 0.144 | 0.144 | 0.273 | 0.097 | 0.048 | 0.021 | 0.029 | 0.016 | 0.009 |
| 9 | 29.8 | 0.132 | 0.142 | 0.273 | 0.101 | 0.050 | 0.021 | 0.030 | 0.016 | 0.009 |
| 10 | 35.4 | 0.120 | 0.141 | 0.273 | 0.105 | 0.051 | 0.021 | 0.030 | 0.017 | 0.009 |
| 11 | 39.3 | 0.110 | 0.141 | 0.273 | 0.108 | 0.052 | 0.021 | 0.031 | 0.017 | 0.009 |
| 12 | 44.6 | 0.102 | 0.142 | 0.273 | 0.110 | 0.053 | 0.021 | 0.032 | 0.017 | 0.009 |
| 13 | 48.7 | 0.094 | 0.144 | 0.273 | 0.112 | 0.053 | 0.020 | 0.032 | 0.017 | 0.009 |
| 14 | 53.7 | 0.087 | 0.148 | 0.273 | 0.114 | 0.052 | 0.020 | 0.033 | 0.018 | 0.009 |
| 15 | 57.0 | 0.081 | 0.152 | 0.273 | 0.115 | 0.052 | 0.019 | 0.034 | 0.018 | 0.009 |
| 16 | 57.5 | 0.076 | 0.158 | 0.273 | 0.115 | 0.051 | 0.019 | 0.035 | 0.018 | 0.009 |
| 17 | 58.9 | 0.073 | 0.165 | 0.273 | 0.115 | 0.049 | 0.018 | 0.035 | 0.019 | 0.009 |
| 18 | 56.8 | 0.070 | 0.173 | 0.273 | 0.114 | 0.048 | 0.017 | 0.036 | 0.019 | 0.009 |

**Table A3:** Relative center of mass position (orange) and radius of gyration in the transverse axis (green) for selected body segments.

| Sex/Age | Center of mass position | | | | | Radius of gyration (transverse axis) | | | | | |
|---|---|---|---|---|---|---|---|---|---|---|---|
| | UpLeg | LowLeg | UpperArm | Forearm | Hand | HeadNeck | LowTrunkPelvis | UpLeg | UpperArm | Forearm | Hand |
| **Male** | | | | | | | | | | | |
| 6 | 0.461 | 0.432 | 0.442 | 0.427 | 0.409 | 0.308 | 0.346 | 0.291 | 0.313 | 0.289 | 0.239 |
| 7 | 0.461 | 0.430 | 0.442 | 0.426 | 0.409 | 0.308 | 0.346 | 0.291 | 0.312 | 0.288 | 0.239 |
| 8 | 0.461 | 0.427 | 0.442 | 0.425 | 0.409 | 0.308 | 0.346 | 0.291 | 0.310 | 0.287 | 0.239 |
| 9 | 0.461 | 0.425 | 0.442 | 0.424 | 0.409 | 0.308 | 0.346 | 0.291 | 0.309 | 0.286 | 0.239 |
| 10 | 0.461 | 0.423 | 0.442 | 0.423 | 0.409 | 0.308 | 0.346 | 0.291 | 0.308 | 0.285 | 0.239 |
| 11 | 0.461 | 0.421 | 0.442 | 0.422 | 0.409 | 0.308 | 0.346 | 0.291 | 0.307 | 0.285 | 0.239 |
| 12 | 0.461 | 0.419 | 0.442 | 0.421 | 0.409 | 0.308 | 0.346 | 0.291 | 0.306 | 0.284 | 0.239 |
| 13 | 0.461 | 0.417 | 0.442 | 0.420 | 0.409 | 0.308 | 0.346 | 0.291 | 0.305 | 0.283 | 0.239 |
| 14 | 0.461 | 0.415 | 0.442 | 0.419 | 0.409 | 0.308 | 0.346 | 0.291 | 0.304 | 0.282 | 0.239 |
| 15 | 0.461 | 0.413 | 0.442 | 0.418 | 0.409 | 0.308 | 0.346 | 0.291 | 0.303 | 0.281 | 0.239 |
| 16 | 0.461 | 0.411 | 0.442 | 0.417 | 0.409 | 0.308 | 0.346 | 0.291 | 0.302 | 0.280 | 0.239 |
| 17 | 0.461 | 0.408 | 0.442 | 0.416 | 0.409 | 0.308 | 0.346 | 0.291 | 0.300 | 0.279 | 0.239 |
| 18 | 0.461 | 0.406 | 0.442 | 0.416 | 0.409 | 0.308 | 0.346 | 0.291 | 0.299 | 0.279 | 0.239 |
| **Female** | | | | | | | | | | | |
| 6 | N/A | N/A | N/A | N/A | N/A | N/A | N/A | N/A | N/A | N/A | N/A |
| 7 | N/A | N/A | N/A | N/A | N/A | N/A | N/A | N/A | N/A | N/A | N/A |
| 8 | N/A | N/A | N/A | N/A | N/A | N/A | N/A | N/A | N/A | N/A | N/A |
| 9 | N/A | N/A | N/A | N/A | N/A | N/A | N/A | N/A | N/A | N/A | N/A |
| 10 | N/A | N/A | N/A | N/A | N/A | N/A | N/A | N/A | N/A | N/A | N/A |
| 11 | N/A | N/A | N/A | N/A | N/A | N/A | N/A | N/A | N/A | N/A | N/A |
| 12 | N/A | N/A | N/A | N/A | N/A | N/A | N/A | N/A | N/A | N/A | N/A |
| 13 | N/A | N/A | N/A | N/A | N/A | N/A | N/A | N/A | N/A | N/A | N/A |
| 14 | N/A | N/A | N/A | N/A | N/A | N/A | N/A | N/A | N/A | N/A | N/A |
| 15 | N/A | N/A | N/A | N/A | N/A | N/A | N/A | N/A | N/A | N/A | N/A |
| 16 | N/A | N/A | N/A | N/A | N/A | N/A | N/A | N/A | N/A | N/A | N/A |
| 17 | N/A | N/A | N/A | N/A | N/A | N/A | N/A | N/A | N/A | N/A | N/A |
| 18 | N/A | N/A | N/A | N/A | N/A | N/A | N/A | N/A | N/A | N/A | N/A |

**Table A4:** Segmental angles (orange) as well as pelvis tilt, pelvic incidence and sacral slope (green).

| Age | Segmental angles | | | | | | | | | | | | | | | | | Pelvis | | |
|---|---|---|---|---|---|---|---|---|---|---|---|---|---|---|---|---|---|---|---|---|
| | L5/S1 | L4/L5 | L3/L4 | L2/L3 | L1/L2 | T12/L1 | T11/T12 | T10/T11 | T9/T10 | T8/T9 | T7/T8 | T6/T7 | T5/T6 | T4/T5 | T3/T4 | T2/T3 | T1/T2 | PT | PI | SS |
| 6 | 23.0 | 13.4 | 8.4 | 4.4 | 1.4 | 0.2 | -1.2 | -2.2 | -2.2 | -4.2 | -5.2 | -4.2 | -5.2 | -4.2 | -4.2 | -3.2 | -1.2 | 4.5 | 43.9 | 39.4 |
| 7 | 24.0 | 15.4 | 8.4 | 5.4 | 1.4 | 0.2 | -1.2 | -2.2 | -2.2 | -4.2 | -5.2 | -5.2 | -5.2 | -4.2 | -4.2 | -3.2 | -1.2 | 5.0 | 44.4 | 39.4 |
| 8 | 24.0 | 15.4 | 8.4 | 5.4 | 1.4 | 0.2 | -1.2 | -2.2 | -2.2 | -4.2 | -5.2 | -5.2 | -5.2 | -4.2 | -4.2 | -3.2 | -1.2 | 5.5 | 45.0 | 39.5 |
| 9 | 24.0 | 15.4 | 8.4 | 5.4 | 1.4 | 0.2 | -1.2 | -2.2 | -2.2 | -4.2 | -5.2 | -5.2 | -5.2 | -4.2 | -4.2 | -3.2 | -1.2 | 6.0 | 45.5 | 39.5 |
| 10 | 26.0 | 17.2 | 9.2 | 6.2 | 2.2 | -1.0 | -1.2 | -2.2 | -3.2 | -4.2 | -5.2 | -6.2 | -5.2 | -4.2 | -4.2 | -3.2 | -1.2 | 6.5 | 47.0 | 40.5 |
| 11 | 26.0 | 17.2 | 9.2 | 6.2 | 2.2 | -1.0 | -1.2 | -2.2 | -3.2 | -4.2 | -5.2 | -6.2 | -5.2 | -4.2 | -4.2 | -3.2 | -1.2 | 7.0 | 47.7 | 40.7 |
| 12 | 26.0 | 17.2 | 9.2 | 6.2 | 2.2 | -1.0 | -1.2 | -2.2 | -3.2 | -4.2 | -5.2 | -6.2 | -5.2 | -4.2 | -4.2 | -3.2 | -1.2 | 7.5 | 48.5 | 41.0 |
| 13 | 24.6 | 16.6 | 9.9 | 6.2 | 2.6 | -1.0 | -1.2 | -2.2 | -3.1 | -4.1 | -6.1 | -5.1 | -6.1 | -5.1 | -4.1 | -3.1 | -3.1 | 8.0 | 49.2 | 41.2 |
| 14 | 24.6 | 16.6 | 9.9 | 6.2 | 2.6 | -1.0 | -1.2 | -2.2 | -3.1 | -4.1 | -6.1 | -5.1 | -6.1 | -5.1 | -4.1 | -3.1 | -3.1 | 8.5 | 50.0 | 41.5 |
| 15 | 24.6 | 16.6 | 9.9 | 6.2 | 2.6 | -1.0 | -1.2 | -2.2 | -3.1 | -4.1 | -6.1 | -5.1 | -6.1 | -5.1 | -4.1 | -3.1 | -3.1 | 9.0 | 50.7 | 41.7 |
| 16 | 24.6 | 16.6 | 9.9 | 6.2 | 2.6 | -1.0 | -1.2 | -2.2 | -3.1 | -4.1 | -6.1 | -5.1 | -6.1 | -5.1 | -4.1 | -3.1 | -3.1 | 9.5 | 51.5 | 42.0 |
| 17 | 24.6 | 16.6 | 9.9 | 6.2 | 2.6 | -1.0 | -1.2 | -2.2 | -3.1 | -4.1 | -6.1 | -5.1 | -6.1 | -5.1 | -4.1 | -3.1 | -3.1 | 10.0 | 52.2 | 42.2 |
| 18 | 24.6 | 16.6 | 9.9 | 6.2 | 2.6 | -1.0 | -1.2 | -2.2 | -3.1 | -4.1 | -6.1 | -5.1 | -6.1 | -5.1 | -4.1 | -3.1 | -3.1 | 10.5 | 52.9 | 42.4 |

**Table A5:** Orange: Cross-sectional areas (CSA [cm$^2$]) for the muscles erector spinae (ES), multifidi (MF), psoas major (PM) and quadratus lumborum (QL) for the levels L3-L5. Green: Scaling factors for all the other levels of ES, MF, PM and QL as well as the remaining trunk muscles rectus abdominis (RA), internal obliques (IO), external obliques (EO), latissimus dorsi (LD) and trapezius (Trap) and the muscles of the upper and lower extremities and the neck (Overall).

| Sex/Age | ES | | | MF | | | PM | | | QL | | Scaling factors | | | | | | |
|---|---|---|---|---|---|---|---|---|---|---|---|---|---|---|---|---|---|---|
| | L3 | L4 | L5 | L3 | L4 | L5 | L3 | L4 | L5 | L3 | L4 | Overall | ES | MF | PM | QL | RA,IO,EO | LD,Trap |
| **Male** | | | | | | | | | | | | | | | | | | |
| 6 | 6.07 | 4.62 | 1.66 | 2.52 | 3.44 | 3.18 | 3.71 | 5.37 | 5.46 | 1.87 | 2.79 | 0.36 | 0.34 | 0.36 | 0.34 | 0.40 | 0.34 | 0.37 |
| 7 | 7.19 | 5.87 | 1.84 | 2.89 | 3.97 | 4.56 | 4.37 | 6.33 | 6.21 | 2.20 | 3.16 | 0.42 | 0.40 | 0.42 | 0.40 | 0.46 | 0.40 | 0.43 |
| 8 | 8.25 | 6.70 | 2.03 | 3.21 | 4.46 | 5.14 | 3.76 | 5.43 | 5.16 | 2.48 | 3.47 | 0.43 | 0.46 | 0.47 | 0.34 | 0.51 | 0.34 | 0.48 |
| 9 | 9.44 | 7.60 | 2.20 | 3.57 | 4.98 | 5.77 | 4.04 | 5.84 | 5.38 | 2.82 | 3.84 | 0.48 | 0.52 | 0.52 | 0.36 | 0.58 | 0.36 | 0.54 |
| 10 | 10.69 | 7.02 | 2.40 | 3.95 | 5.52 | 5.16 | 5.61 | 8.09 | 7.25 | 3.22 | 4.26 | 0.57 | 0.59 | 0.57 | 0.50 | 0.64 | 0.50 | 0.60 |
| 11 | 11.79 | 7.72 | 2.56 | 4.26 | 5.97 | 5.62 | 6.95 | 10.03 | 8.80 | 3.56 | 4.62 | 0.64 | 0.64 | 0.62 | 0.61 | 0.71 | 0.61 | 0.66 |
| 12 | 12.97 | 8.44 | 2.70 | 4.60 | 6.46 | 5.85 | 7.67 | 11.07 | 9.56 | 3.94 | 5.06 | 0.70 | 0.71 | 0.67 | 0.67 | 0.78 | 0.67 | 0.72 |
| 13 | 14.15 | 13.90 | 2.78 | 4.93 | 6.91 | 8.11 | 7.99 | 11.57 | 9.92 | 4.37 | 5.53 | 0.75 | 0.77 | 0.72 | 0.69 | 0.86 | 0.69 | 0.78 |
| 14 | 15.29 | 14.94 | 2.90 | 5.26 | 7.36 | 8.68 | 8.66 | 12.57 | 10.70 | 4.81 | 6.05 | 0.81 | 0.83 | 0.77 | 0.75 | 0.94 | 0.75 | 0.84 |
| 15 | 16.31 | 15.87 | 3.02 | 5.56 | 7.76 | 9.20 | 9.94 | 14.46 | 12.29 | 5.23 | 6.58 | 0.89 | 0.88 | 0.81 | 0.86 | 1.02 | 0.86 | 0.90 |
| 16 | 17.23 | 16.70 | 3.13 | 5.83 | 8.12 | 9.68 | 10.62 | 15.50 | 13.21 | 5.67 | 7.16 | 0.95 | 0.93 | 0.85 | 0.92 | 1.11 | 0.92 | 0.96 |
| 17 | 18.20 | 17.54 | 3.23 | 6.13 | 8.48 | 10.18 | 11.42 | 16.73 | 14.37 | 6.18 | 7.86 | 1.02 | 0.98 | 0.89 | 1.00 | 1.22 | 1.00 | 1.03 |
| 18 | 18.88 | 18.13 | 3.33 | 6.35 | 8.74 | 10.57 | 12.05 | 17.72 | 15.41 | 6.63 | 8.54 | 1.07 | 1.01 | 0.92 | 1.06 | 1.31 | 1.06 | 1.08 |
| **Female** | | | | | | | | | | | | | | | | | | |
| 6 | 4.14 | 3.64 | 1.43 | 1.61 | 2.38 | 2.15 | 2.90 | 4.21 | 4.28 | 1.09 | 1.77 | 0.38 | 0.32 | 0.32 | 0.46 | 0.37 | 0.46 | 0.34 |
| 7 | 4.59 | 3.96 | 1.58 | 1.85 | 2.74 | 3.07 | 3.07 | 4.45 | 4.37 | 1.29 | 2.01 | 0.41 | 0.35 | 0.37 | 0.47 | 0.42 | 0.47 | 0.38 |
| 8 | 4.69 | 4.12 | 1.74 | 1.87 | 2.76 | 3.25 | 3.55 | 5.13 | 4.87 | 1.22 | 1.93 | 0.43 | 0.36 | 0.37 | 0.54 | 0.40 | 0.54 | 0.38 |
| 9 | 5.37 | 4.67 | 1.89 | 2.09 | 3.09 | 3.65 | 3.50 | 5.06 | 4.70 | 1.38 | 2.14 | 0.46 | 0.42 | 0.42 | 0.53 | 0.45 | 0.53 | 0.43 |
| 10 | 6.10 | 5.22 | 2.08 | 2.30 | 3.42 | 4.06 | 3.94 | 5.68 | 5.12 | 1.58 | 2.38 | 0.52 | 0.47 | 0.46 | 0.59 | 0.51 | 0.59 | 0.48 |
| 11 | 6.71 | 5.73 | 2.22 | 2.48 | 3.70 | 4.42 | 4.32 | 6.24 | 5.51 | 1.74 | 2.57 | 0.56 | 0.51 | 0.49 | 0.65 | 0.55 | 0.65 | 0.52 |
| 12 | 7.38 | 6.27 | 2.35 | 2.68 | 4.00 | 4.79 | 4.77 | 6.89 | 5.99 | 1.93 | 2.81 | 0.62 | 0.56 | 0.53 | 0.71 | 0.61 | 0.71 | 0.57 |
| 13 | 8.55 | 7.44 | 2.39 | 2.87 | 4.27 | 5.12 | 5.29 | 7.66 | 6.61 | 2.14 | 3.08 | 0.67 | 0.61 | 0.57 | 0.78 | 0.67 | 0.78 | 0.62 |
| 14 | 9.24 | 8.00 | 2.50 | 3.07 | 4.56 | 5.48 | 5.76 | 8.36 | 7.16 | 2.35 | 3.37 | 0.73 | 0.66 | 0.61 | 0.85 | 0.74 | 0.85 | 0.67 |
| 15 | 9.29 | 8.50 | 2.60 | 3.24 | 4.80 | 5.81 | 6.19 | 9.02 | 7.71 | 2.56 | 3.67 | 0.78 | 0.70 | 0.64 | 0.91 | 0.80 | 0.91 | 0.72 |
| 16 | 9.81 | 8.95 | 2.70 | 3.40 | 5.02 | 6.12 | 6.62 | 9.66 | 8.29 | 2.78 | 3.99 | 0.83 | 0.74 | 0.68 | 0.98 | 0.87 | 0.98 | 0.76 |
| 17 | 10.39 | 9.42 | 2.78 | 3.58 | 5.26 | 6.44 | 7.11 | 10.43 | 9.01 | 3.04 | 4.39 | 0.90 | 0.78 | 0.71 | 1.06 | 0.96 | 1.06 | 0.82 |
| 18 | 10.77 | 9.73 | 2.86 | 3.71 | 5.42 | 6.69 | 7.50 | 11.04 | 9.66 | 3.26 | 4.77 | 0.95 | 0.81 | 0.73 | 1.12 | 1.03 | 1.12 | 0.86 |

**Table A6:** Lower extremity, pelvis and lumbar spine joint angles used for the simulations in the validation studies.

| | *pelvis_tilt* | *hip_flexion_r* | *hip_flexion_l* | *Abs_LB* | *Abs_AR* | *L5_S1_FE* | *L5_S1_LB* | *L5_S1_AR* | *L4_L5_FE* | *L4_L5_LB* | *L4_L5_AR* | *L3_L4_FE* | *L3_L4_LB* | *L3_L4_AR* | *L2_L3_FE* | *L2_L3_LB* | *L2_L3_AR* | *L1_L2_FE* | *L1_L2_LB* | *L1_L2_AR* |
|---|---|---|---|---|---|---|---|---|---|---|---|---|---|---|---|---|---|---|---|---|
| **Andersen, Peltonen** | | | | | | | | | | | | | | | | | | | | |
| Upright standing | 0 | 0 | 0 | 0 | 0 | 0 | 0 | 0 | 0 | 0 | 0 | 0 | 0 | 0 | 0 | 0 | 0 | 0 | 0 | 0 |
| **Sinaki** | | | | | | | | | | | | | | | | | | | | |
| Prone, 6-9 year olds | -80 | -10 | -10 | 0 | 0 | 0 | 0 | 0 | 0 | 0 | 0 | 0 | 0 | 0 | 0 | 0 | 0 | 0 | 0 | 0 |
| Prone, 10-18 year olds | -80 | -10 | -10 | 0 | 0 | 0 | 0 | 0 | 0 | 0 | 0 | 0 | 0 | 0 | 0 | 0 | 0 | 0 | 0 | 0 |
| Supine, 6-9 year olds | 90 | 0 | 0 | 0 | 0 | 0 | 0 | 0 | 0 | 0 | 0 | 0 | 0 | 0 | 0 | 0 | 0 | 0 | 0 | 0 |
| Supine, 10-18 year olds | 85 | 5 | 5 | 0 | 0 | 0 | 0 | 0 | 0 | 0 | 0 | 0 | 0 | 0 | 0 | 0 | 0 | 0 | 0 | 0 |
| **Neuschwander** | | | | | | | | | | | | | | | | | | | | |
| Upright standing | 0 | 0 | 0 | 0 | 0 | 0 | 0 | 0 | 0 | 0 | 0 | 0 | 0 | 0 | 0 | 0 | 0 | 0 | 0 | 0 |
| **Schultz** | | | | | | | | | | | | | | | | | | | | |
| Upright standing | 0 | 0 | 0 | 0 | 0 | 0 | 0 | 0 | 0 | 0 | 0 | 0 | 0 | 0 | 0 | 0 | 0 | 0 | 0 | 0 |
| Hands on chest | 0 | 0 | 0 | 0 | 0 | 0 | 0 | 0 | 0 | 0 | 0 | 0 | 0 | 0 | 0 | 0 | 0 | 0 | 0 | 0 |
| Hands in front | 0 | 0 | 0 | 0 | 0 | 0 | 0 | 0 | 0 | 0 | 0 | 0 | 0 | 0 | 0 | 0 | 0 | 0 | 0 | 0 |
| Hip flexion, hands in front | -30 | 30 | 30 | 0 | 0 | 0 | 0 | 0 | 0 | 0 | 0 | 0 | 0 | 0 | 0 | 0 | 0 | 0 | 0 | 0 |
| **Polga** | | | | | | | | | | | | | | | | | | | | |
| Upright standing | 0 | 0 | 0 | 0 | 0 | 0 | 0 | 0 | 0 | 0 | 0 | 0 | 0 | 0 | 0 | 0 | 0 | 0 | 0 | 0 |
| Extension | 0 | 0 | 0 | 0 | 0 | 0.514 | 0 | 0 | 1.117 | 0 | 0 | 1.599 | 0 | 0 | 1.873 | 0 | 0 | 2.281 | 0 | 0 |
| Lateral bending | 0 | 0 | 0 | 5.420 | 0 | 0 | 0.734 | 0 | 0 | 0.982 | 0 | 0 | 1.330 | 0 | 0 | 1.356 | 0 | 0 | 1.018 | 0 |
| Flexion | -30 | 30 | 30 | 0 | 0 | 0 | 0 | 0 | 0 | 0 | 0 | 0 | 0 | 0 | 0 | 0 | 0 | 0 | 0 | 0 |
| Axial rotation (twisting) | 0 | 0 | 0 | 0 | 5.136 | 0 | 0 | 1.068 | 0 | 0 | 1.134 | 0 | 0 | 1.134 | 0 | 0 | 0.933 | 0 | 0 | 0.867 |
| Elbows flexed | 0 | 0 | 0 | 0 | 0 | 0 | 0 | 0 | 0 | 0 | 0 | 0 | 0 | 0 | 0 | 0 | 0 | 0 | 0 | 0 |

**Table A7:** Thoracic spine (T6-T12) joint angles used for the simulations in the validation studies.

| | T12_L1_FE | T12_L1_LB | T12_L1_AR | T11_T12_FE | T11_T12_LB | T11_T12_AR | T10_T11_FE | T10_T11_LB | T10_T11_AR | T9_T10_FE | T9_T10_LB | T9_T10_AR | T8_T9_FE | T8_T9_LB | T8_T9_AR | T7_T8_FE | T7_T8_LB | T7_T8_AR | T6_T7_FE | T6_T7_LB | T6_T7_AR |
|---|---|---|---|---|---|---|---|---|---|---|---|---|---|---|---|---|---|---|---|---|---|
| **Andersen, Peltonen** | | | | | | | | | | | | | | | | | | | | | |
| Upright standing | 0 | 0 | 0 | 0 | 0 | 0 | 0 | 0 | 0 | 0 | 0 | 0 | 0 | 0 | 0 | 0 | 0 | 0 | 0 | 0 | 0 |
| **Sinaki** | | | | | | | | | | | | | | | | | | | | | |
| Prone, 6-9 year olds | 0 | 0 | 0 | 0 | 0 | 0 | 0 | 0 | 0 | 0 | 0 | 0 | 0 | 0 | 0 | 0 | 0 | 0 | 0 | 0 | 0 |
| Prone, 10-18 year olds | 0 | 0 | 0 | 0 | 0 | 0 | 0 | 0 | 0 | 0 | 0 | 0 | 0 | 0 | 0 | 0 | 0 | 0 | 0 | 0 | 0 |
| Supine, 6-9 year olds | 0 | 0 | 0 | 0 | 0 | 0 | 0 | 0 | 0 | 0 | 0 | 0 | 0 | 0 | 0 | 0 | 0 | 0 | 0 | 0 | 0 |
| Supine, 10-18 year olds | 0 | 0 | 0 | 0 | 0 | 0 | 0 | 0 | 0 | 0 | 0 | 0 | 0 | 0 | 0 | 0 | 0 | 0 | 0 | 0 | 0 |
| **Neuschwander** | | | | | | | | | | | | | | | | | | | | | |
| Upright standing | 0 | 0 | 0 | 0 | 0 | 0 | 0 | 0 | 0 | 0 | 0 | 0 | 0 | 0 | 0 | 0 | 0 | 0 | 0 | 0 | 0 |
| **Schultz** | | | | | | | | | | | | | | | | | | | | | |
| Upright standing | 0 | 0 | 0 | 0 | 0 | 0 | 0 | 0 | 0 | 0 | 0 | 0 | 0 | 0 | 0 | 0 | 0 | 0 | 0 | 0 | 0 |
| Hands on chest | 0 | 0 | 0 | 0 | 0 | 0 | 0 | 0 | 0 | 0 | 0 | 0 | 0 | 0 | 0 | 0 | 0 | 0 | 0 | 0 | 0 |
| Hands in front | 0 | 0 | 0 | 0 | 0 | 0 | 0 | 0 | 0 | 0 | 0 | 0 | 0 | 0 | 0 | 0 | 0 | 0 | 0 | 0 | 0 |
| Hip flexion, hands in front | 0 | 0 | 0 | 0 | 0 | 0 | 0 | 0 | 0 | 0 | 0 | 0 | 0 | 0 | 0 | 0 | 0 | 0 | 0 | 0 | 0 |
| **Polga** | | | | | | | | | | | | | | | | | | | | | |
| Upright standing | 0 | 0 | 0 | 0 | 0 | 0 | 0 | 0 | 0 | 0 | 0 | 0 | 0 | 0 | 0 | 0 | 0 | 0 | 0 | 0 | 0 |
| Extension | 1.169 | 0 | 0 | 1.169 | 0 | 0 | 0.876 | 0 | 0 | 0.584 | 0 | 0 | 0.584 | 0 | 0 | 0.584 | 0 | 0 | 0.487 | 0 | 0 |
| Lateral bending | 0 | 1.802 | 0 | 0 | 1.884 | 0 | 0 | 1.474 | 0 | 0 | 1.310 | 0 | 0 | 1.064 | 0 | 0 | 1.392 | 0 | 0 | 0.900 | 0 |
| Flexion | 0 | 0 | 0 | 0 | 0 | 0 | 0 | 0 | 0 | 0 | 0 | 0 | 0 | 0 | 0 | 0 | 0 | 0 | 0 | 0 | 0 |
| Axial rotation (twisting) | 0 | 0 | 0.582 | 0 | 0 | 1.509 | 0 | 0 | 3.021 | 0 | 0 | 3.138 | 0 | 0 | 2.904 | 0 | 0 | 2.673 | 0 | 0 | 2.208 |
| Elbows flexed | 0 | 0 | 0 | 0 | 0 | 0 | 0 | 0 | 0 | 0 | 0 | 0 | 0 | 0 | 0 | 0 | 0 | 0 | 0 | 0 | 0 |

**Table A8:** Thoracic spine (T1-T6), head/neck and upper extremity joint angles used for the simulations in the validation studies.

| | T5_T6_FE | T5_T6_LB | T5_T6_AR | T4_T5_FE | T4_T5_LB | T4_T5_AR | T3_T4_FE | T3_T4_LB | T3_T4_AR | T2_T3_FE | T2_T3_LB | T2_T3_AR | T1_T2_FE | T1_T2_LB | T1_T2_AR | T1_head_neck_FE | shoulder_elv_r | shoulder_rot_r | elv_angle_r | elbow_flexion_r | shoulder_elv_l | shoulder_rot_l | elv_angle_l | elbow_flexion_l |
|---|---|---|---|---|---|---|---|---|---|---|---|---|---|---|---|---|---|---|---|---|---|---|---|---|
| **Andersen, Peltonen** | | | | | | | | | | | | | | | | | | | | | | | | |
| Upright standing | 0 | 0 | 0 | 0 | 0 | 0 | 0 | 0 | 0 | 0 | 0 | 0 | 0 | 0 | 0 | 0 | 0 | 0 | 0 | 0 | 0 | 0 | 0 | 0 |
| **Sinaki** | | | | | | | | | | | | | | | | | | | | | | | | |
| Prone, 6-9 year olds | 0 | 0 | 0 | 0 | 0 | 0 | 0 | 0 | 0 | 0 | 0 | 0 | 0 | 0 | 0 | 15 | 0 | 0 | -10 | 0 | 0 | 0 | -10 | 0 |
| Prone, 10-18 year olds | 0 | 0 | 0 | 0 | 0 | 0 | 0 | 0 | 0 | 0 | 0 | 0 | 0 | 0 | 0 | 15 | 0 | 0 | -10 | 0 | 0 | 0 | -10 | 0 |
| Supine, 6-9 year olds | 0 | 0 | 0 | 0 | 0 | 0 | 0 | 0 | 0 | 0 | 0 | 0 | 0 | 0 | 0 | -10 | 0 | 0 | 0 | 0 | 0 | 0 | 0 | 0 |
| Supine, 10-18 year olds | 0 | 0 | 0 | 0 | 0 | 0 | 0 | 0 | 0 | 0 | 0 | 0 | 0 | 0 | 0 | -10 | 0 | 0 | 5 | 0 | 0 | 0 | 5 | 0 |
| **Neuschwander** | | | | | | | | | | | | | | | | | | | | | | | | |
| Upright standing | 0 | 0 | 0 | 0 | 0 | 0 | 0 | 0 | 0 | 0 | 0 | 0 | 0 | 0 | 0 | 0 | 0 | 0 | 0 | 0 | 0 | 0 | 0 | 0 |
| **Schultz** | | | | | | | | | | | | | | | | | | | | | | | | |
| Upright standing | 0 | 0 | 0 | 0 | 0 | 0 | 0 | 0 | 0 | 0 | 0 | 0 | 0 | 0 | 0 | 0 | 0 | 0 | 0 | 0 | 0 | 0 | 0 | 0 |
| Hands on chest | 0 | 0 | 0 | 0 | 0 | 0 | 0 | 0 | 0 | 0 | 0 | 0 | 0 | 0 | 0 | 0 | 0 | 44 | 15 | 140 | 0 | -44 | 15 | 140 |
| Hands in front | 0 | 0 | 0 | 0 | 0 | 0 | 0 | 0 | 0 | 0 | 0 | 0 | 0 | 0 | 0 | 0 | 0 | 20 | 80 | 0 | 0 | -20 | 80 | 0 |
| Hip flexion, hands in front | 0 | 0 | 0 | 0 | 0 | 0 | 0 | 0 | 0 | 0 | 0 | 0 | 0 | 0 | 0 | 0 | 0 | 0 | 110 | 0 | 0 | 0 | 110 | 0 |
| **Polga** | | | | | | | | | | | | | | | | | | | | | | | | |
| Upright standing | 0 | 0 | 0 | 0 | 0 | 0 | 0 | 0 | 0 | 0 | 0 | 0 | 0 | 0 | 0 | 0 | 0 | 0 | 0 | 0 | 0 | 0 | 0 | 0 |
| Extension | 0.389 | 0 | 0 | 0.389 | 0 | 0 | 0.389 | 0 | 0 | 0.389 | 0 | 0 | 0.389 | 0 | 0 | 0 | 0 | 0 | 0 | 0 | 0 | 0 | 0 | 0 |
| Lateral bending | 0 | 0.656 | 0 | 0 | 0.738 | 0 | 0 | 1.146 | 0 | 0 | 1.064 | 0 | 0 | 1.146 | 0 | 0 | 20 | 0 | 0 | 0 | 0 | 0 | 0 | 0 |
| Flexion | 0 | 0 | 0 | 0 | 0 | 0 | 0 | 0 | 0 | 0 | 0 | 0 | 0 | 0 | 0 | 0 | 0 | 0 | 30 | 0 | 0 | 0 | 30 | 0 |
| Axial rotation (twisting) | 0 | 0 | 2.091 | 0 | 0 | 1.860 | 0 | 0 | 1.626 | 0 | 0 | 1.860 | 0 | 0 | 1.395 | 0 | 0 | 0 | 0 | 0 | 0 | 0 | 0 | 0 |
| Elbows flexed | 0 | 0 | 0 | 0 | 0 | 0 | 0 | 0 | 0 | 0 | 0 | 0 | 0 | 0 | 0 | 0 | 0 | 0 | 0 | 90 | 0 | 0 | 0 | 90 |

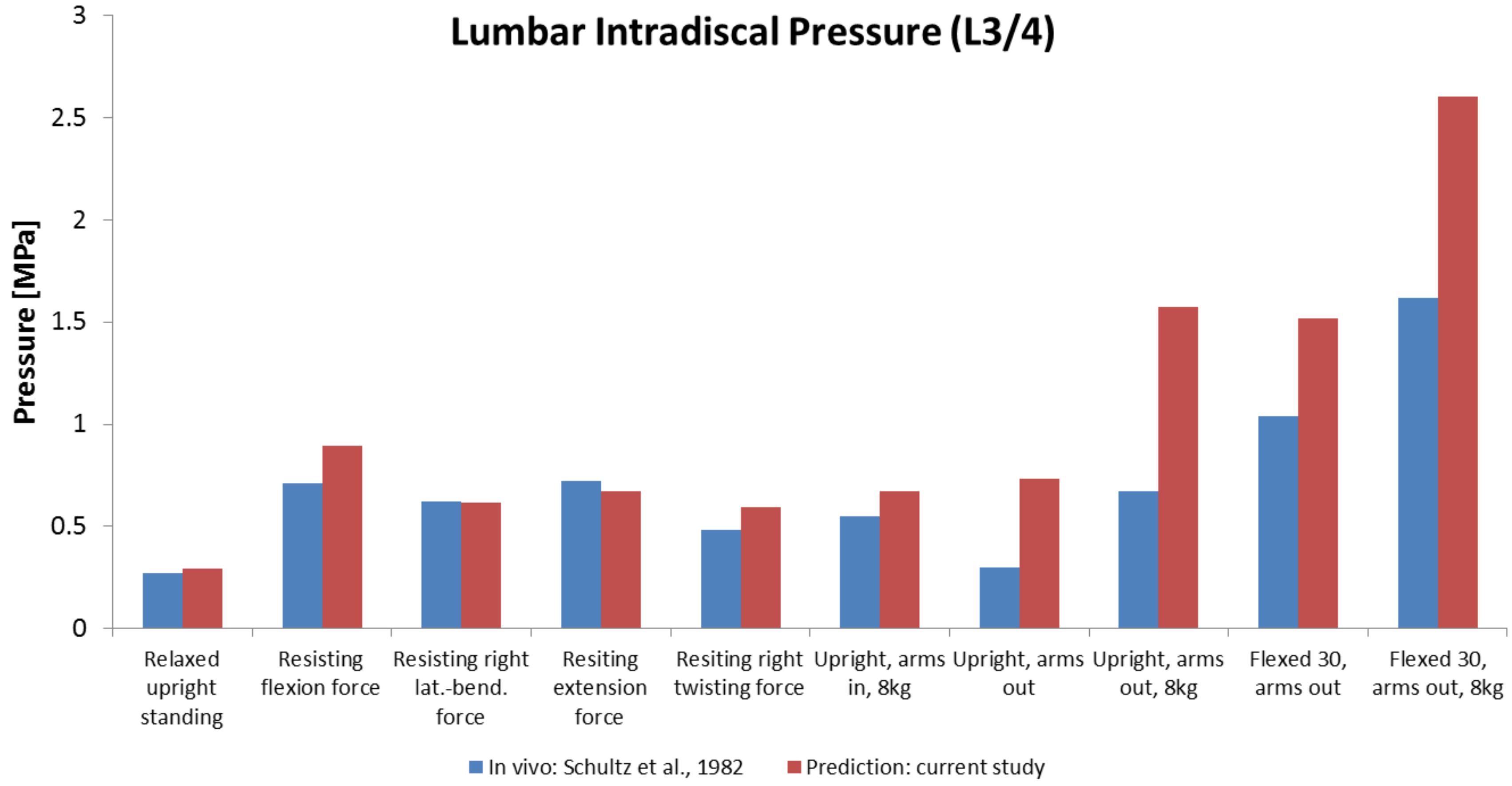

**Figure A1:** Lumbar intradiscal pressure (L3/4 level) when resisting forces on the thorax and in different upper body positions with and without carrying weights. In vivo data were retrieved from Schultz et al., 1982 for young adults (mean age: 21.8 years) having a transducer inserted in the third lumbar disc.

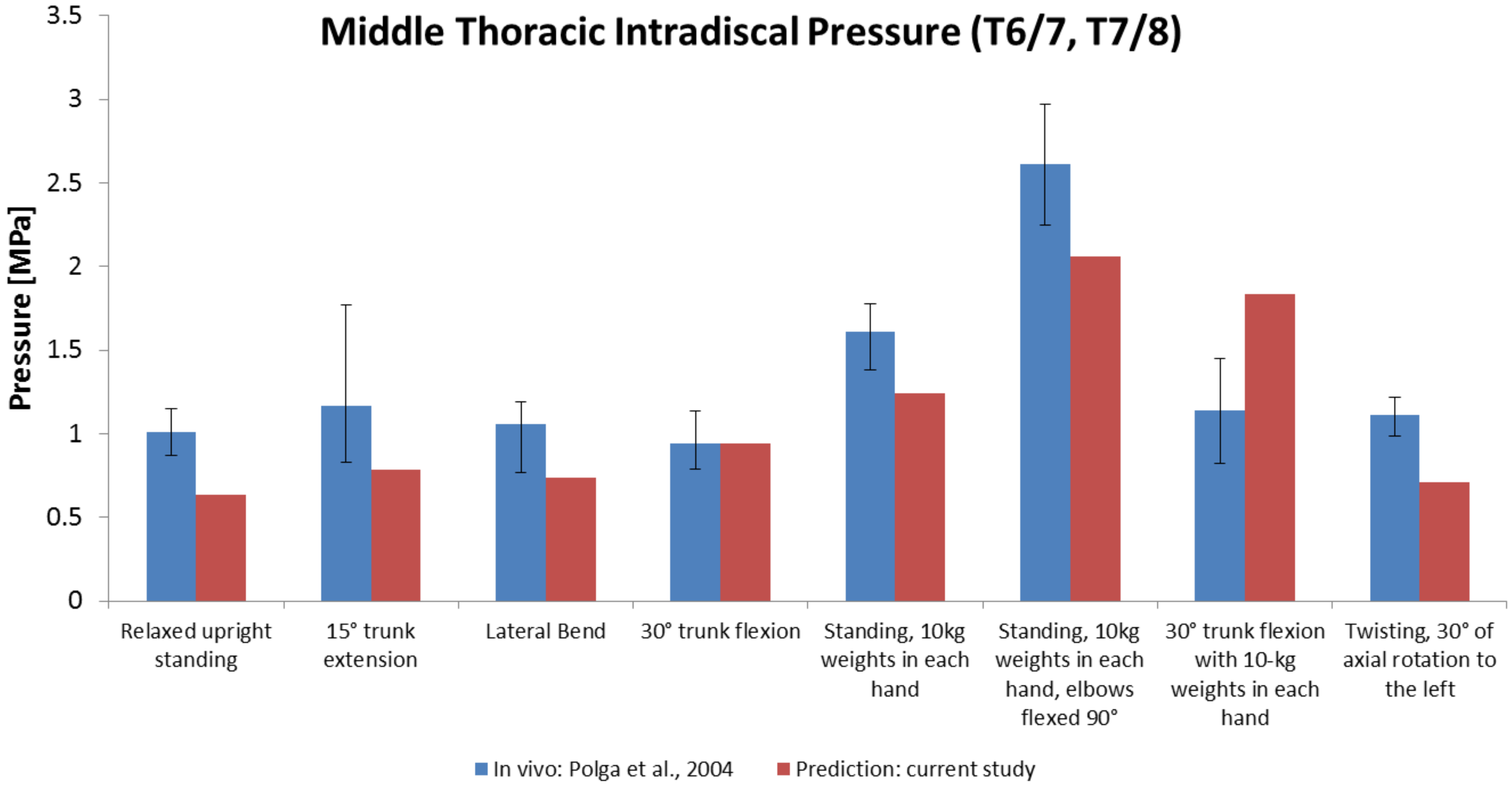

**Figure A2:** Middle thoracic intradiscal pressure (average of T6/7 and T7/8 levels) in different upper body positions with and without carrying weights. In vivo data were retrieved from Polga et al., 2004 for adults aged 19-47 years having a transducer inserted in the respective thoracic discs. Error bars indicate the range of values in the in vivo study.

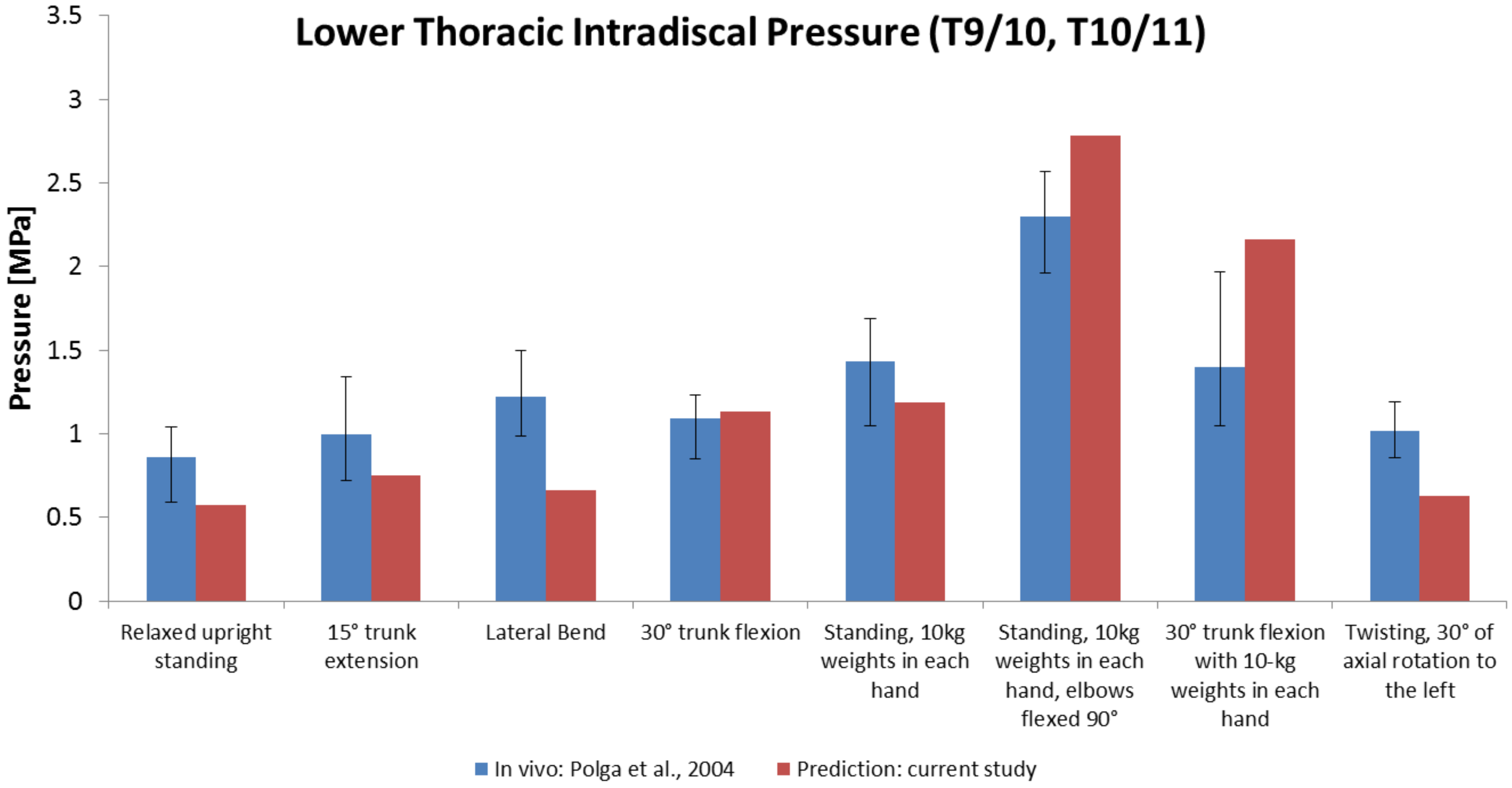

**Figure A3:** Lower thoracic intradiscal pressure (average of T9/10 and T10/11 levels) in different upper body positions with and without carrying weights. In vivo data were retrieved from Polga et al., 2004 for adults aged 19-47 years having a transducer inserted in the respective thoracic discs. Error bars indicate the range of values in the in vivo study.

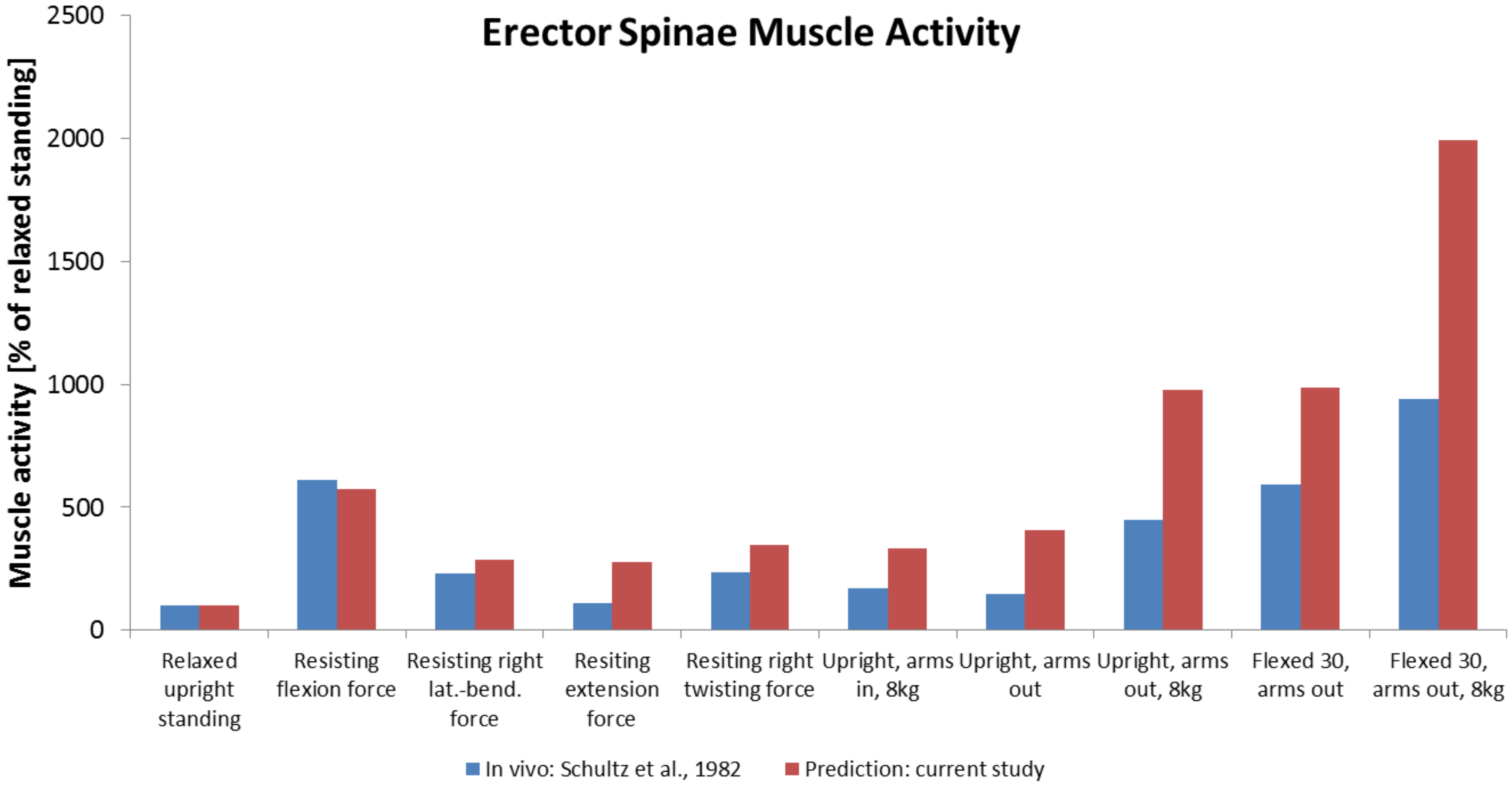

**Figure A4:** Erector spinae muscle activity when resisting forces on the thorax and in different upper body positions with and without carrying weights. In vivo data were retrieved from Schultz et al., 1982 for young adults (mean age: 21.8 years) having a transducer inserted in the third lumbar disc.

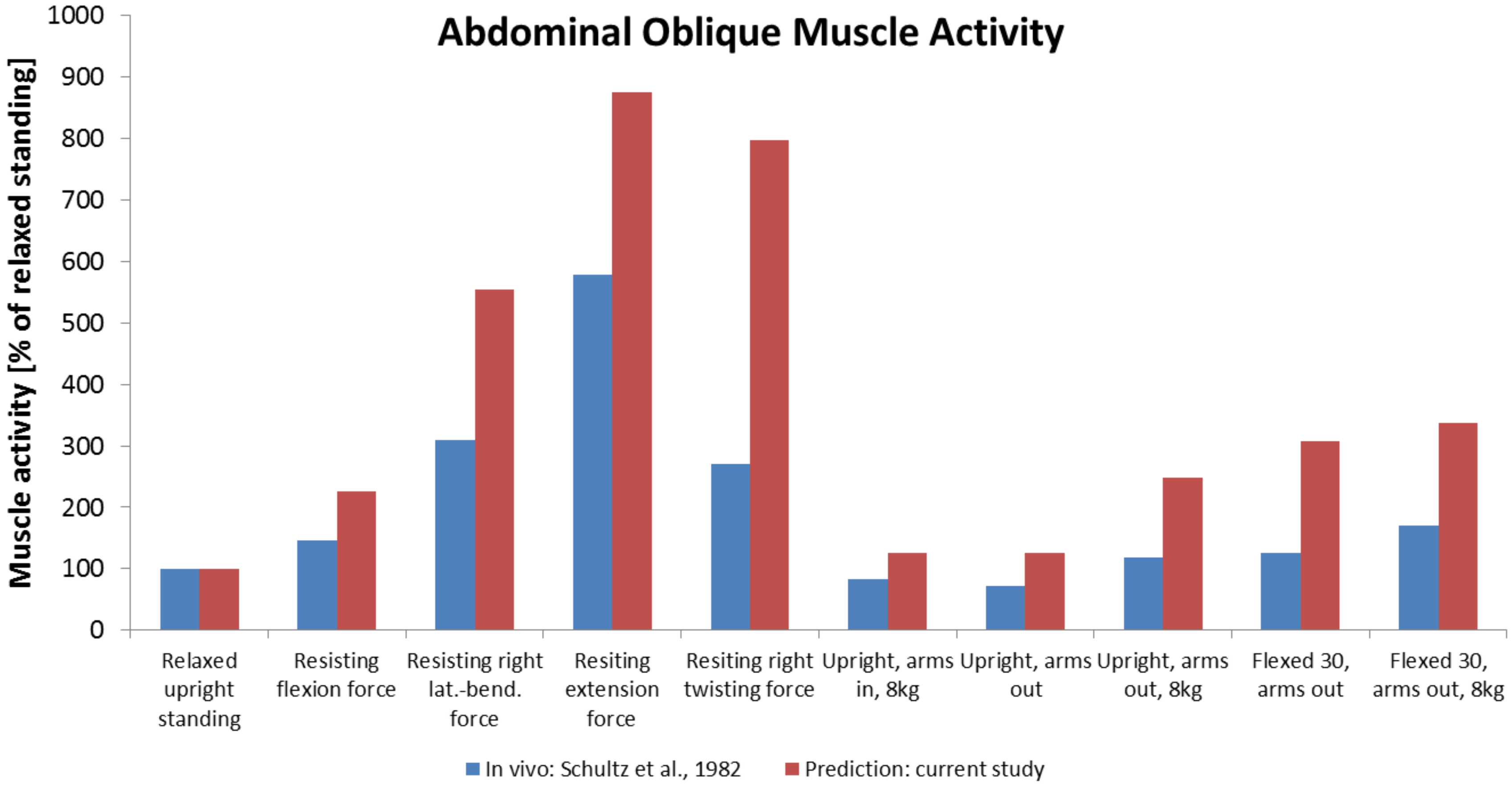

**Figure A5:** Abdominal oblique muscle activity when resisting forces on the thorax and in different upper body positions with and without carrying weights. In vivo data were retrieved from Schultz et al., 1982 for young adults (mean age: 21.8 years) having a transducer inserted in the third lumbar disc.